\documentclass[twocolumn]{aastex701}

\usepackage{bm}
\usepackage{amsmath}
\usepackage{comment}

\graphicspath{{./}{figures/}}

\begin{document}

\title{Revisiting the Photometric Modulation Amplitudes of Kepler Planet Hosts:\\A Confounder-Adjusted Test of Stellar Spin--Orbit Alignment}

\author[0000-0003-1298-9699]{Kento Masuda}
\affiliation{Department of Earth and Space Science, Graduate School of Science, Osaka University, Osaka, Japan}
\email{kmasuda@ess.sci.osaka-u.ac.jp}

\author[0000-0002-1027-0990]{Subo Dong}
\affiliation{Department of Astronomy, School of Physics, Peking University, Beijing, China}
\affiliation{Kavli Institute of Astronomy and Astrophysics, Peking University, Beijing, China}
\affiliation{National Astronomical Observatories, Chinese Academy of Sciences, Beijing, China}
\email{dongsubo@pku.edu.cn}

\newcommand{\koi}{\mathrm{KOI}}
\newcommand{\kic}{\mathrm{KIC}}
\newcommand{\kicrs}{\mathrm{KIC}^*}

\newcommand{\tra}{\mathcal{T}} 
\renewcommand{\det}{\mathcal{D}} 
\newcommand{\rot}{\mathcal{R}} 
\newcommand{\X}{\mathcal{X}}
\newcommand{\prot}{\log P_\mathrm{rot}}
\newcommand{\amp}{r} 
\newcommand{\teff}{T_\mathrm{eff}} 
\newcommand{\teffi}{T_{\mathrm{eff},i}} 
\renewcommand{\mag}{m_\mathrm{Kep}} 
\newcommand{\rad}{\log R_\star}

\newcommand{\pdet}{\pi_\mathrm{D|TR}} 
\newcommand{\pstar}{\theta_\ast}
\newcommand{\pplanet}{\theta_{\rm p}}
\newcommand{\z}{z}
\renewcommand{\d}{\mathrm{d}}
\newcommand{\mazeh}{2015ApJ...801....3M}
\defcitealias{\mazeh}{M15}

\begin{abstract}

T. Mazeh and coworkers pioneered a method using rotational photometric modulation as a proxy to infer stellar obliquities in Kepler transiting planetary systems, based on the expected positive correlation between spot-induced modulation amplitudes and stellar inclinations. They found that planet hosts cooler than the Sun showed larger modulation amplitudes than comparison stars, consistent with low stellar obliquities. For hotter hosts, however, they reported the opposite --- a negative amplitude difference (hosts minus comparison stars) --- which is difficult to interpret in terms of viewing geometry. We show that this comparison is affected by confounding variables not controlled for in the sample selection. One such confounder is apparent magnitude, whose distributions differ markedly between the two samples for hotter stars and affect the observed modulation amplitudes. We develop resampling and likelihood-based approaches that enable comparison of the amplitude distributions at fixed apparent magnitude, stellar radius, and rotation period. The previously reported negative amplitude difference for hot hosts is no longer present in either analysis. The likelihood-based analysis finds a positive amplitude difference up to $\sim 6100\,\mathrm{K}$ that becomes consistent with zero at higher temperatures, while the resampling analysis shows a similar but less pronounced trend. This result is qualitatively consistent with independent $v\sin i$ analyses suggesting low obliquities for cooler super-Earth and sub-Neptune hosts and broader obliquity distributions for hotter ones. This may also indicate a gradual change in the obliquity distribution with stellar temperature, although changes in spot distributions or selection effects could also contribute to the photometric trends in hot stars.

\end{abstract}



\section{Introduction}\label{sec:intro}

Rossiter--McLaughlin (RM) measurements of hot Jupiters have shown that close-in giant planets span a wide range of spin--orbit configurations \citep[e.g.,][]{2018haex.bookE...2T,2022PASP..134h2001A}.
These measurements also revealed a clear dependence on host-star temperature: hot Jupiters around stars cooler than 
the Kraft break tend to have low obliquities, whereas those around hotter stars show a broader distribution \citep{2010ApJ...718L.145W,2010ApJ...719..602S}. 

One interpretation of this temperature trend is that the low obliquities of cool stars are due to tidal realignment of the stellar spin \citep{2010ApJ...718L.145W,2012ApJ...757...18A}, in which case the present-day alignment of cool hosts need not be primordial and hot Jupiters in general could initially have a broad obliquity distribution. 
Such a distribution could arise naturally from high-eccentricity migration, rather than from smooth migration within an initially aligned disk  \citep[see review by][]{2018ARA&A..56..175D}.
Alternatively, the trend may arise from properties of the protoplanetary disks rather than from the planets themselves. For example, disks may frequently form misaligned with their host stars and/or subsequently undergo realignment in a manner that depends on stellar temperature or mass \citep[e.g.,][]{2012Natur.491..418B,2014MNRAS.440.3532L,2015MNRAS.450.3306F,2015ApJ...811...82S, biddle_one-third_2025}. In this case, even hot Jupiters that migrate smoothly within their disks could end up on misaligned orbits.
These possibilities are not mutually exclusive and may operate simultaneously. 
In any case, the temperature dependence provides an important clue to how stellar spins and planetary orbital planes are established and subsequently evolve.

Obliquity measurements for close-in super-Earths and sub-Neptunes with radii $\lesssim 4\,R_\oplus$,\footnote{This radius threshold is motivated by the observation that larger planets appear to constitute a distinct population, with lower occurrence rates and a preference for metal-rich hosts \citep{2018AJ....155...89P}. Their elevated orbital eccentricities \citep{2025PNAS..12205295G} also suggest that they may not represent a dynamically unperturbed population.} which have been discovered in large numbers by transit surveys such as Kepler \citep{2010Sci...327..977B} and TESS \citep{2015JATIS...1a4003R}, help distinguish between these interpretations. 
Their typical masses are roughly an order of magnitude or more lower than those of hot Jupiters, resulting in much weaker tidal interactions.
The stellar obliquities in these systems are therefore expected to retain more primordial information. 
Many of these systems also contain multiple transiting planets, suggesting that nearly coplanar configurations are common \citep{2011ApJS..197....8L,2012AJ....143...94T,2014ApJ...790..146F}; the intrinsic fraction of such systems may be even higher after accounting for transit geometry and detection incompleteness \citep{2018ApJ...860..101Z,2019MNRAS.490.4575H}. Such near-coplanar architectures also suggest that many of these systems have preserved orbital configurations inherited from their protoplanetary disks.

\begin{figure}
    \centering
    \includegraphics[width=\columnwidth,clip]{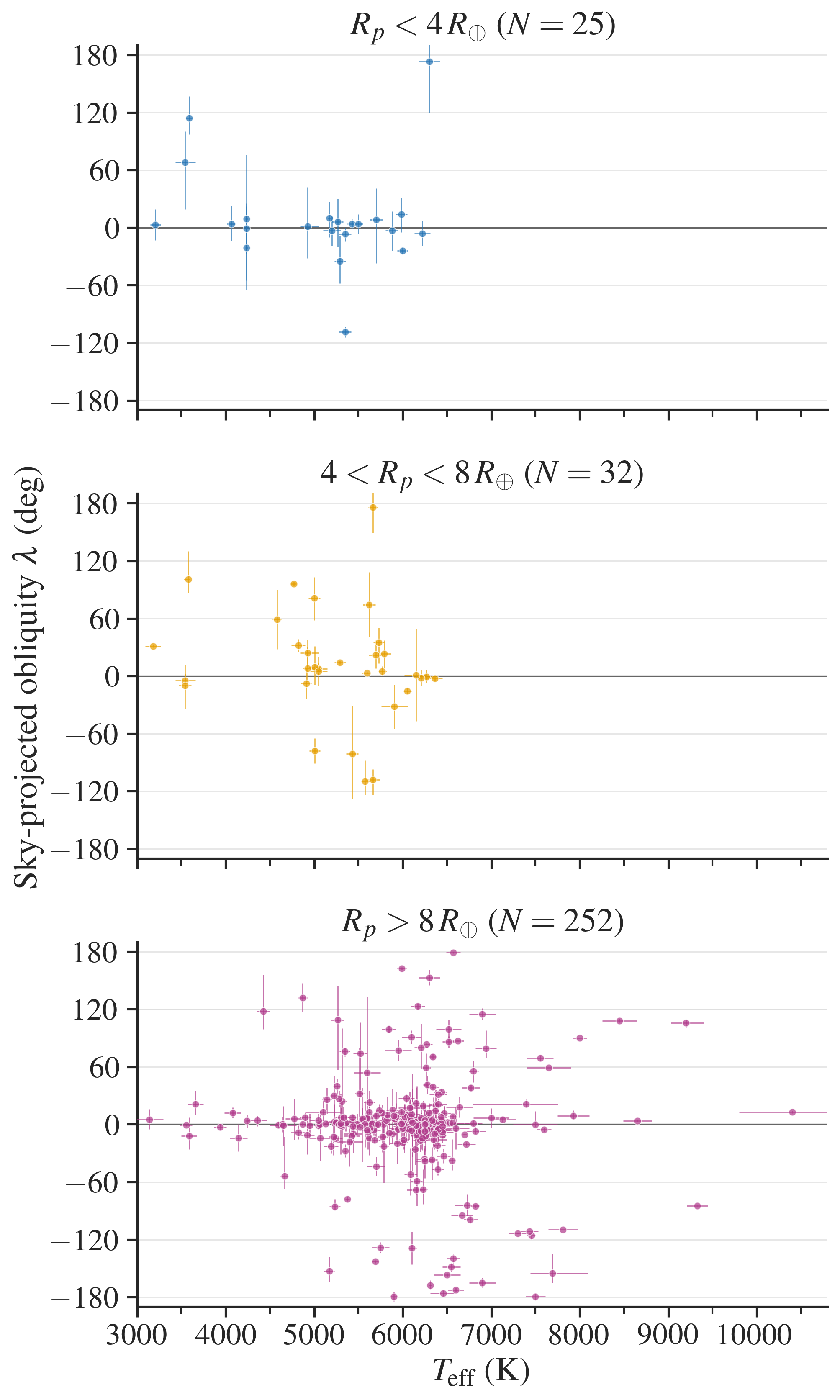}
    \caption{Sky-projected stellar obliquity $\lambda$ as a function of host-star effective temperature for planets with orbital periods $P<100\,\mathrm{days}$. From top to bottom, the panels show systems with $R_p<4\,R_\oplus$, $4\,R_\oplus<R_p<8\,R_\oplus$, and $R_p>8\,R_\oplus$. The obliquities, stellar effective temperatures, orbital periods, and published planetary radii are compiled from TEPCat \citep{2026Univ...12...62S}; the horizontal and vertical error bars show the cataloged uncertainties in $T_\mathrm{eff}$ and $\lambda$, respectively, where available. This is a heterogeneous compilation rather than a uniformly selected sample, and the measurements differ in precision and reliability. Accordingly, this figure is intended to illustrate the current state of the available measurements rather than to support a statistical inference about the underlying obliquity distributions. The giant-planet sample spans a wide range of obliquities but also shows a pronounced concentration of aligned systems near $\lambda=0^\circ$, extending up to $T_\mathrm{eff}\sim 6500\,\mathrm{K}$ \citep{2026ApJ...996L...7W}.}
    \label{fig:rm}
\end{figure}

Despite the observational challenges compared with hot Jupiters, the number of RM measurements for such small transiting planets has been growing in recent years (Figure~\ref{fig:rm}). 
Although a population analysis based on a well-defined selection function has yet to be performed, the existing measurements suggest that low obliquities might be common among small-planet systems around cool stars (top panel). At least for systems with multiple transiting planets, statistically significant evidence for preferentially low obliquities has been reported \citep{2013ApJ...771...11A,2024A&A...690A.379K,2025AJ....170..182P,2026arXiv260323713H}. However, the current measurements are restricted to stars cooler than $\sim 6200\,\mathrm{K}$.

Utilizing a large sample of transiting planets for which orbits are known to be edge-on, population-level studies focusing on spin-axis inclinations have also been performed \citep[e.g.,][]{2010ApJ...719..602S,2012ApJ...756...66H,2016ApJ...819...85C,2017AJ....154..270W,2018AJ....156..253M,2021AJ....161...68L}.
In particular, a comparison of the projected rotation velocity $v\sin i$ with the equatorial rotation velocity $v$ inferred from the stellar rotation period and radius suggests low obliquities for small planets around cool stars \citep{2017AJ....154..270W,2018AJ....156..253M}.
Using a $v\sin i$-only comparison likely less sensitive to rotation-period selection, \citet{2021AJ....161...68L} further supported this result, finding larger mean $v\sin i$ for cool transiting-planet hosts ($5950$--$6250\,\mathrm{K}$) than for matched control stars.
Although the sample of hotter stars in \citet{2021AJ....161...68L} was smaller, both that study and a subsequent larger-sample analysis \citep{2024ApJ...968L...2L} reported evidence for a broader obliquity distribution at higher temperatures. This apparent similarity to the hot-Jupiter trend suggests that the temperature dependence may not arise solely from tidal realignment induced by hot Jupiters.

Here we focus on another statistical approach based on the amplitudes of rotational photometric modulation, which has the potential to probe stellar obliquities of small planets across a broad range of stellar temperatures. The method was pioneered by \citet[][hereafter M15]{2015ApJ...801....3M}, who compared the rotational modulation amplitudes of Kepler stars with detected transiting planets (Kepler Objects of Interest; KOIs) with those of a reference sample of Kepler stars (KIC stars).
This comparison probes the correlation between the stellar spin and planetary orbit: if they are aligned, stars with transiting planets tend to be viewed edge-on, resulting in larger modulation amplitudes if spots tend to occur at low latitudes, as on the Sun.
For cool stars, they found that planet hosts showed stronger modulation, consistent with low obliquities. For hotter stars, however, they found the opposite trend: planet hosts showed smaller modulation amplitudes than those in the comparison sample. Although this trend may be partly affected by transit-detectability bias, the hot-star result is not straightforward to interpret in terms of viewing geometry alone \citep{2015ApJ...801....3M,2016ApJ...818....5L}.

We revisit this problem because comparisons of photometric amplitudes may be affected by confounding variables that were not explicitly controlled for in \citetalias{\mazeh}. Here, confounding variables refer to stellar properties correlated with both the modulation amplitude and whether a star belongs to the planet-host or control sample. If the amplitude depends on these properties and their distributions differ between the two samples, the observed amplitude distributions may differ even if the underlying obliquity distributions are identical. 
In Section~\ref{sec:concept}, we explain this concept and identify the additional variables that should be controlled. Section~\ref{sec:principle} then gives a unified mathematical treatment of these confounding variables and the relevant selection effects. 
In Section~\ref{sec:resampling}, we present a resampling-based analysis that reweights and resamples the control stars to match the stellar-parameter distribution of the planet-host sample, thereby mitigating the effects of confounding variables. In Section~\ref{sec:flow}, we present a likelihood-based analysis that explicitly models the joint distribution of variability amplitude and stellar properties, allowing the amplitude distributions to be compared at fixed stellar properties.
In Section~\ref{sec:teff_robustness}, we repeat the analysis using a smaller subset of stars with effective temperatures from LAMOST low-resolution spectroscopy and find no evidence that the results are sensitive to the source of the effective temperatures.
Both analyses show that controlling for confounding variables removes the previously reported negative amplitude offset for hotter KOIs. The likelihood-based analysis yields a positive median shift of approximately $0.07$--$0.09\,\mathrm{dex}$ below $\sim 5{,}500\,\mathrm{K}$; the shift decreases with increasing temperature, remains positive up to $\sim 6{,}100\,\mathrm{K}$, and is consistent with zero at higher temperatures. The resampling analysis yields a qualitatively consistent but less pronounced trend. We interpret these results and discuss limitations of this type of analysis in Section~\ref{sec:discussion}, and summarize our main findings in Section~\ref{sec:summary}.

\subsection{Summary of the Symbols}

Table~\ref{tab:symbols} summarizes the variables and conditions (selection criteria) discussed in this paper. For conditions, the notation $\mathcal{AB}$ denotes the logical conjunction of $\mathcal{A}$ and $\mathcal{B}$. The photometric modulation amplitude, rotation period, and stellar radius are expressed as base-10 logarithms.

\begin{table}[ht]
\centering
\caption{Summary of the Symbols}\label{tab:symbols}
\begin{tabular}{ll}
\hline\hline
Symbol & Meaning\\
\hline
$\rot$ & \parbox[t]{0.72\columnwidth}{KIC star with rotational modulation detected in \cite{\mazeh} or in \citet{2014ApJS..211...24M}}\\
$\tra$ & star with a transiting planet\\
$\det$ & \parbox[t]{0.72\columnwidth}{star with a detected transiting planet (always conditioned on $\tra$)}\\
$\koi=\det\tra\rot$ & KOI with detected rotational modulation\\
$\amp=\log R$ & \parbox[t]{0.72\columnwidth}{logarithmic amplitude of rotational modulation (ppm)}\\
$\teff$ & effective temperature of a star (K)\\
$\mag$ & Kepler magnitude of a star\\
$\prot$ & logarithmic rotation period of a star (day)\\
$\rad$ & logarithmic radius of a star ($R_\odot$)\\
\hline
\end{tabular}
\end{table}

\section{Why Matching in Effective Temperature Is Not Enough}\label{sec:concept}

\citetalias{\mazeh} compared the photometric modulation amplitudes $\amp$ of KOIs with those measured for KIC stars by \citet{2014ApJS..211...24M} using the same method, to investigate the correlation between stellar spins and planetary orbits. Because transiting planets have nearly edge-on orbits, such a correlation would produce different stellar inclination distributions for KOIs and KIC stars, potentially leading to different amplitude distributions.
\citetalias{\mazeh} performed this comparison 
as a function of effective temperature $\teff$. This is a natural comparison given the temperature dependence of stellar obliquities observed in hot-Jupiter systems. Comparing stars at fixed $\teff$ also helps control for differences in the temperature distributions of the two samples, which can affect their amplitude distributions because $\amp$ is correlated with $\teff$.

From this latter perspective, however, fixing $\teff$ alone is not sufficient.
Stars have other properties, some of which are correlated with $\amp$. In the comparison performed by \citetalias{\mazeh} at fixed $\teff$, these properties are marginalized over. If such variables have different distributions among KIC and KOI stars, the amplitude distributions will differ because of these variables, not because of inclinations. For this to occur, the correlation between a ``confounder'' and the amplitude need not be causal or intrinsic; a statistical correlation introduced by, e.g., selection criteria is sufficient.
\citetalias{\mazeh} did examine one such effect: because rotation period is correlated with modulation amplitude, they tried controlling for the rotation period $\prot$ within relatively broad ($\sim 500\,\mathrm{K}$) $\teff$ bins, and found that this did not substantially change their result.

\begin{figure*}[ht]
    \plotone{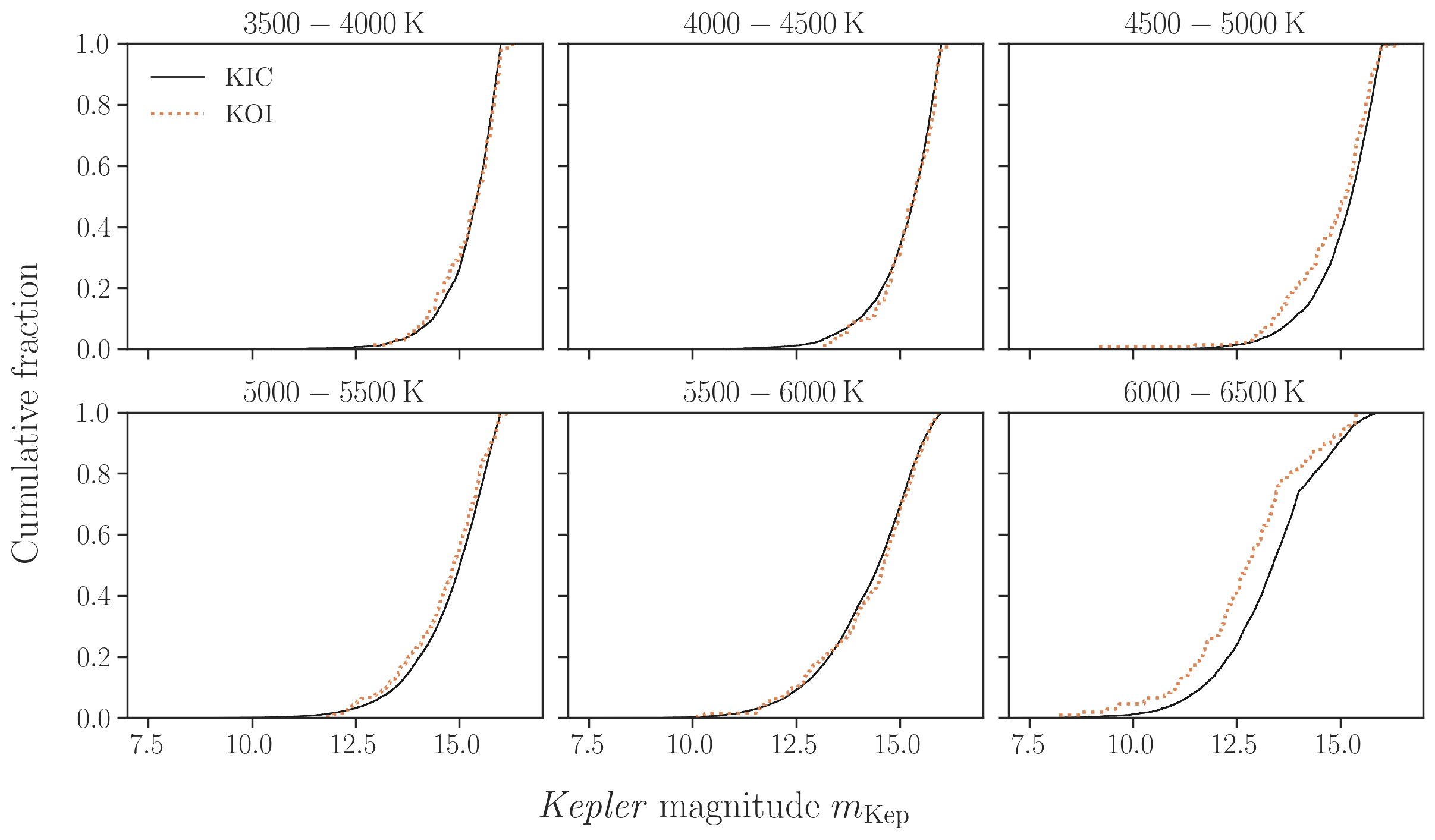}
    \caption{The Kepler magnitude distributions of KIC (black solid) and KOI (orange dotted) stars in the \citetalias{\mazeh} sample.}
    \label{fig:cdf_kepmag}
\end{figure*}

One important variable that was not controlled in \citetalias{\mazeh} is the apparent Kepler magnitude $\mag$. Although it is not obvious whether photometric amplitudes should {\it intrinsically} depend on $\mag$, it certainly does when we consider stars {\it with detected rotational modulation} in \citet{2014ApJS..211...24M} or in \citetalias{\mazeh}, as the weakest modulations are missed among the faintest stars \citep{2022ApJ...933..195M,2022ApJ...937...94M}. As shown in Figure~\ref{fig:cdf_kepmag}, the $\mag$ distributions in their KIC and KOI samples differ, especially for hotter stars, where KOIs are on average brighter. Thus, weaker modulation can be detected among the brighter KOIs than among the fainter KIC stars, and the $\amp$ distribution of KOIs is biased toward smaller values when marginalized over $\mag$. This aligns with the sign of the KOI--KIC amplitude difference reported by \citetalias{\mazeh} for hot stars. See also Section~\ref{sec:resampling} for the differences in the other variables.

Given these considerations, at least two more variables are worth controlling for.
As mentioned above, $\prot$ remains important given its correlation with $\amp$. Its distribution may differ between KICs and KOIs, as $\prot$ is related to stellar activity and hence the noise level in the Kepler light curve that affects planet detectability.
Another good candidate is the stellar radius $\rad$, which was also not considered in \citetalias{\mazeh}. 
For stars somewhat more massive than the Sun, the radius at a given $\teff$ reflects the evolutionary phase and hence the convective-zone depth, which is related to $\amp$ at a given $\prot$ through the associated convective turnover time and hence the Rossby number \citep[e.g.,][]{1984ApJ...279..763N,2021ApJ...912..127S,2021A&A...652L...2C}.
The $\rad$ distributions of KOIs and KIC stars can differ for multiple reasons, given that $\rad$ affects both the geometric transit probability and the transit depth.

Thus, in this work, we control for $\z=(\mag,\rad,\prot)$ in addition to $\teff$. We do so jointly because their correlations with one another and with $\amp$ may also matter. 
It is interesting to note that the distributions of $\mag$ and $\rad$ are broader for stars hotter than the Sun, which span different evolutionary phases and hence a range of sizes and intrinsic brightnesses, so the impact of this correction is likely to be larger for them.
In addition to these confounding issues, we need to account for the fact that the observed KOI sample is shaped by planet detectability, as was also discussed by \cite{\mazeh}, and that both samples are conditioned on the detection of rotational modulation. The next section gives a unified mathematical formulation of these effects.

\section{Formal Treatment of Confounding Variables and Selection Effects}\label{sec:principle}

We now formalize the discussion in Section~\ref{sec:concept} in probabilistic terms, starting with the comparison performed by \citetalias{\mazeh}. We then examine how confounding variables and selection effects enter this comparison, providing the basis for the analyses in Sections~\ref{sec:resampling} and~\ref{sec:flow}.

\subsection{Revisiting the \citetalias{\mazeh} Comparison}

\citet{\mazeh} compared the photometric modulation amplitudes of KIC and KOI stars as a function of $\teff$ to investigate stellar spin--orbit alignment. Their main comparison used $250\,\mathrm{K}$ temperature bins, with the median value of $\amp$ evaluated separately for each sample in each bin.

The binning can be viewed as a way to examine the conditional amplitude distribution $p(\amp|\teff)$ using samples drawn from the joint distribution $p(\amp,\teff)$, with this interpretation becoming exact in the narrow-bin limit.
When we select stars in a bin centered on $\teff=\tau$, we assign weights $w(\teff-\tau)$ to the observed pairs $(\amp_i,\teffi)$ in each sample, with a constant weight inside the bin and zero outside it. 
The resulting amplitude distribution of the selected stars is given by weighting the joint distribution and normalizing over $\amp$:
\begin{align}
\notag
    &{\int p(\amp,\teff)\,w(\teff-\tau)\,\d\teff
    \over \int p(\teff)\,w(\teff-\tau)\,\d\teff}.
\end{align}
In the narrow-bin limit, the normalized weight function approaches $\delta(\teff-\tau)$, so this reduces to
\begin{align}
\notag
    \left.p(\amp,\teff)\over p(\teff)\right|_{\teff=\tau} 
    =\left.p(\amp|\teff)\,p(\teff)\over p(\teff)\right|_{\teff=\tau} 
    = p(\amp|\tau).
\end{align}
In practice, \citetalias{\mazeh} used the median to summarize this conditional amplitude distribution.

Applying this procedure separately to the KIC and KOI samples, we can therefore describe the comparison of \citetalias{\mazeh} as a comparison of (the medians of)
\begin{align}
\notag
    p(\amp|\teff,\X) \quad\mathrm{and}\quad
    p(\amp|\teff,\X\tra),
\end{align}
where $\X$ denotes the common selection criteria and $\tra$ denotes the presence of a transiting planet. This comparison examines how the amplitude distribution differs with $\tra$ at a given $\teff$. For now, this description omits the additional detection requirements of the observed samples; we return to them below.

As discussed in Section~\ref{sec:concept}, examining this difference as a function of $\teff$ is not only physically motivated but also helps distinguish a difference associated with $\tra$ from one caused by different temperature distributions. 
To illustrate this second point, consider what would happen if the two samples were compared without conditioning on $\teff$. The corresponding marginal distributions are
\begin{align}
\notag
    p(\amp|\X) &= \int p(\amp,\teff|\X)\,\d\teff \\
    &= \int p(\amp|\teff,\X)\,p(\teff|\X)\,\d\teff,\quad \notag\\
\notag
    p(\amp|\X\tra) &= \int p(\amp,\teff|\X\tra)\d\teff \\
\notag
    &= \int p(\amp|\teff,\X\tra)\,p(\teff|\X\tra)\,\d\teff
\end{align}
These expressions average the conditional amplitude distributions, $p(\amp|\teff,\X)$ and $p(\amp|\teff,\X\tra)$, over the temperature distribution of each sample, $p(\teff|\X)$ and $p(\teff|\X\tra)$. If $\amp$ depends on $\teff$, differences between $p(\teff|\X)$ and $p(\teff|\X\tra)$ can produce different marginal amplitude distributions even when the conditional distributions at each $\teff$ are identical. Comparing the samples at fixed $\teff$ avoids this source of difference.

In summary, \citetalias{\mazeh} used binning to compare $\amp$ distributions conditioned on $\teff$ and interpreted the KOI--KIC difference as an effect of conditioning on $\tra$.
Below, we clarify two limitations of this comparison: conditioning only on $\teff$ and not explicitly accounting for the detection requirements of the observed samples. 

\subsection{Importance of Controlling Other Variables}\label{ssec:method_control}

We first consider how other stellar properties can affect the comparison, as discussed in Section~\ref{sec:concept}.
Each star has properties other than $\amp$ and $\teff$, which we collectively denote as $\z$. When comparing $p(\amp|\teff)$, these variables are implicitly marginalized over their joint distribution specific to each sample:
\begin{align}
\notag
    p(\amp|\teff,\X) &= \int p(\amp,\z|\teff,\X)\,d\z \\
\notag
    &= \int p(\amp|\z,\teff,\X)\,p(\z|\teff,\X)\,\d\z,
    \quad \\
    p(\amp|\teff,\X\tra) 
    &= \int p(\amp|\z,\teff,\X\tra)\,p(\z|\teff,\X\tra)\,\d\z. 
    \notag
\end{align}

This is the same issue illustrated above for $\teff$: if $\amp$ depends on $\z$ when conditioned on $\teff$ and $\X$ or $\X\tra$, the difference between $p(\z|\teff,\X)$ and $p(\z|\teff,\X\tra)$ results in a difference between $p(\amp|\teff,\X)$ and $p(\amp|\teff,\X\tra)$. We should therefore compare $p(\amp|\z,\teff,\X)$ and $p(\amp|\z,\teff,\X\tra)$, and any such stellar parameter needs to be included in $\z$ on the right side of the bar.
For the present analysis, we adopt $\z=(\mag,\rad,\prot)$, as discussed in Section~\ref{sec:concept}, and compare the amplitude distributions at fixed $(\z,\teff)$ using the resampling and likelihood-based approaches described below.

\subsection{Impacts of Conditioning on Detected Transiting Planets and Rotational Modulation}

Another important consideration is that both the KIC and KOI samples require detected rotational modulation, while the KOI sample additionally requires a detected transiting planet. We therefore need to account for how these detection requirements affect the comparison. Writing the rotational-modulation and planet-detection conditions as $\rot$ and $\det$, respectively, the observed comparison is between $\X\rot$ and $\X\rot\det\tra$, whereas the comparison of interest is between $\X$ and $\X\tra$. For clarity, we suppress $\X$ from the notation below, with all distributions understood to be conditioned on it.

We first consider the impact of $\det$. Since we have
\begin{align}
     \notag
    p(\amp,\z,\teff|\det\tra\rot) &= {p(\det|\amp,\z,\teff,\tra\rot)\,p(\amp,\z,\teff|\tra\rot)\over p(\det|\tra\rot)} \\
    &\equiv  {\pdet(\amp,\z,\teff)\,p(\amp,\z,\teff|\tra\rot) \over p(\det|\tra\rot)},
    \label{eq:pdet}
\end{align}
one can ``correct for'' the $\det$ effect to obtain $p(\amp,\z,\teff|\tra\rot)$ essentially by reweighting the samples from $\det\tra\rot$ by the inverse\footnote{Inverse weighting can be unstable when $\pdet$ is small, and we do not use it in practice below.} of the detectability of a {\it transiting} planet $\pdet(\amp,\z,\teff)\equiv p(\det|\amp,\z,\teff,\tra\rot)$ defined in the last equation.
We assume that this term can be estimated from the available information about the Kepler planet-detection pipeline, and discuss the implementation of this correction in Section~\ref{ssec:resampling_implementation}.

Such a correction essentially allows us to compare the $\tra\rot$ sample against the $\rot$ sample. As we discussed above, we wish to perform this comparison in a manner conditioned on $(\z,\teff)$:
\begin{align}
    \notag
    &{p(\amp|\z,\teff,\tra\rot) \over p(\amp|\z,\teff,\rot)} \\
    &= {p(\rot|\amp,\z,\teff,\tra)\,p(\amp|\z,\teff,\tra) \over p(\rot|\z,\teff,\tra)}
    \left({p(\rot|\amp,\z,\teff)\,p(\amp|\z,\teff) \over p(\rot|\z,\teff)}\right)^{-1} \notag\\
    &\propto {p(\rot|\amp,\z,\teff,\tra) \over p(\rot|\amp,\z,\teff)}{p(\amp|\z,\teff,\tra) \over p(\amp|\z,\teff)} \notag
\end{align}
focusing only on the $\amp$ dependence. 
If we assume that the detectability of rotational modulation $\rot$ is not affected by the presence of transiting planets at fixed $(\amp,\z,\teff)$, i.e., $p(\rot|\amp,\z,\teff,\tra)=p(\rot|\amp,\z,\teff)$, then the first factor in the above equation cancels. The ratio of the two amplitude distributions conditioned on $\rot$ therefore has the same $\amp$ dependence as the corresponding ratio without conditioning on $\rot$. In particular, conditioning on $\rot$  does not by itself generate an amplitude difference under the null hypothesis $p(\amp|\z,\teff,\tra)=p(\amp|\z,\teff)$, which forms the basis of the comparison below.
We assume that this is the case in the samples of \citetalias{\mazeh} and \citet{2014ApJS..211...24M}, given that the same pipeline was applied to the KOI sample after masking transits and that $\rot$ is conditioned not only on $\amp$ but also on the main relevant variables in $\z$, particularly apparent magnitude \citep[see][]{2022ApJ...933..195M,2022ApJ...937...94M}.\footnote{
In general, the condition $\tra$ could affect $\rot$ through stellar inclination: if the stellar spin and planetary orbit tend to be aligned, for example, stars with transiting planets tend to be viewed edge-on, and so may have larger $\amp$. But here we consider the comparison conditioned on $\amp$, $\z$, which should be mostly sufficient for determining the detectability, so this path is effectively accounted for.
}

We reiterate that these arguments rely on \((\z,\teff)\) adequately accounting for the relevant confounding variables, as discussed in Section~\ref{ssec:method_control}. Any unaccounted-for variable that affects the amplitude distribution and is distributed differently between stars with and without transiting planets could still bias the comparison.

\section{Detectability-Weighted Resampling}\label{sec:resampling}

The considerations in Section~\ref{sec:principle} show that the effects of the identified confounding variables and planet detectability can be accounted for by conditioning the comparison on $(\teff,\z)$ and subjecting the KIC sample to the same planet-detection selection as the KOI sample through detectability weighting. Here we implement this comparison by resampling the detectability-weighted KIC sample.\footnote{The code used for the analyses in this paper is available at \url{https://github.com/kemasuda/kepler-phot-var}.} This procedure extends the weighted-sample approach of \citetalias{\mazeh}, which conditioned the comparison on $\teff$ alone, to incorporate planet detectability and the additional conditioning variables $\z$.

\begin{figure*}[ht]
    \plotone{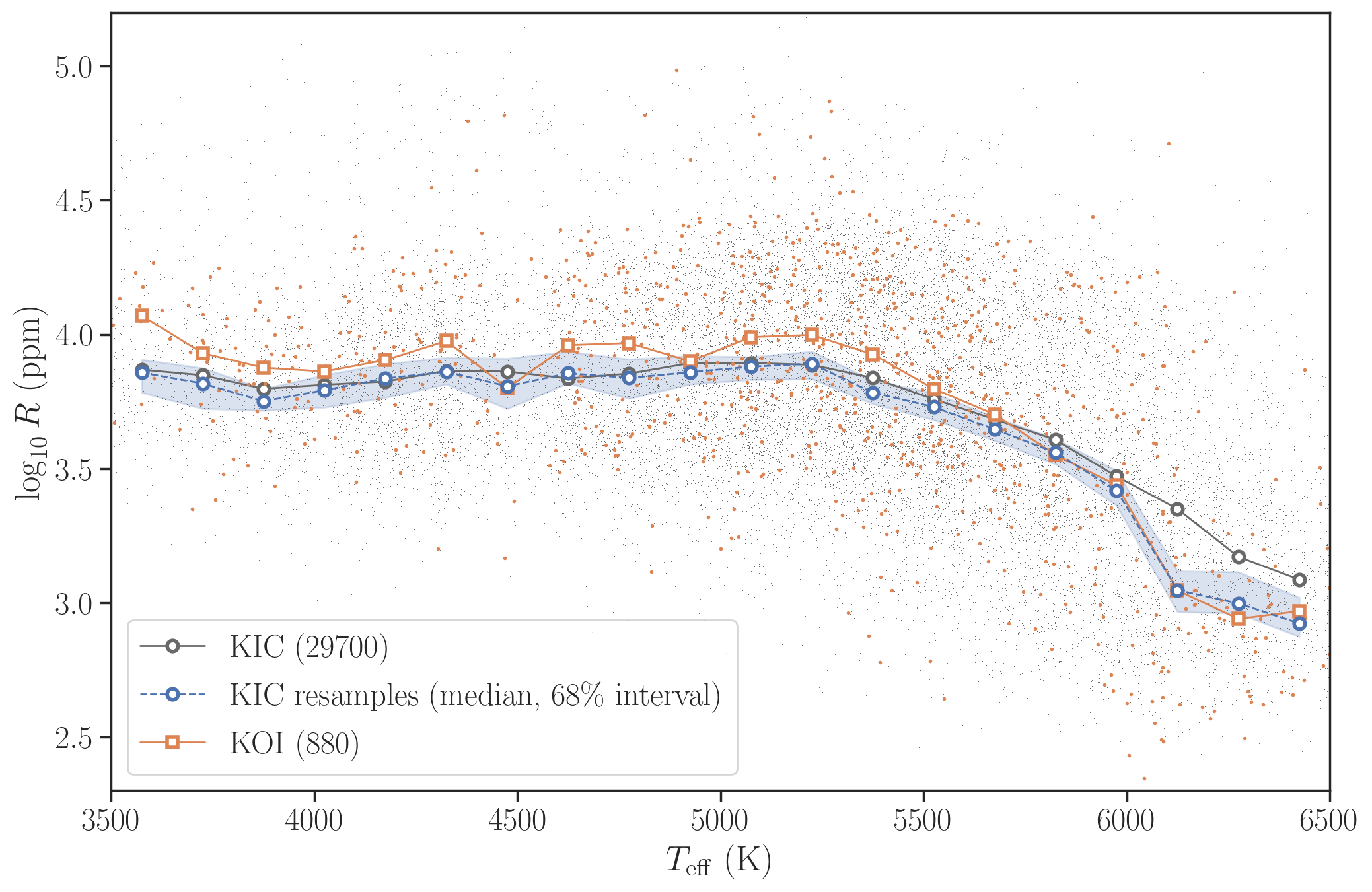}
    \caption{Median photometric modulation amplitude as a function of effective temperature for the original KIC sample (gray circles), the KIC resamples (blue circles), and the KOI sample (orange squares). Small gray and orange points show the individual KIC and KOI stars, respectively. The blue shaded region shows the 16th--84th percentile interval among the 200 resampling realizations. Medians are evaluated in $150\,\mathrm{K}$ bins.}
    \label{fig:resample_m15}
\end{figure*}

\begin{figure*}[ht]
    \plotone{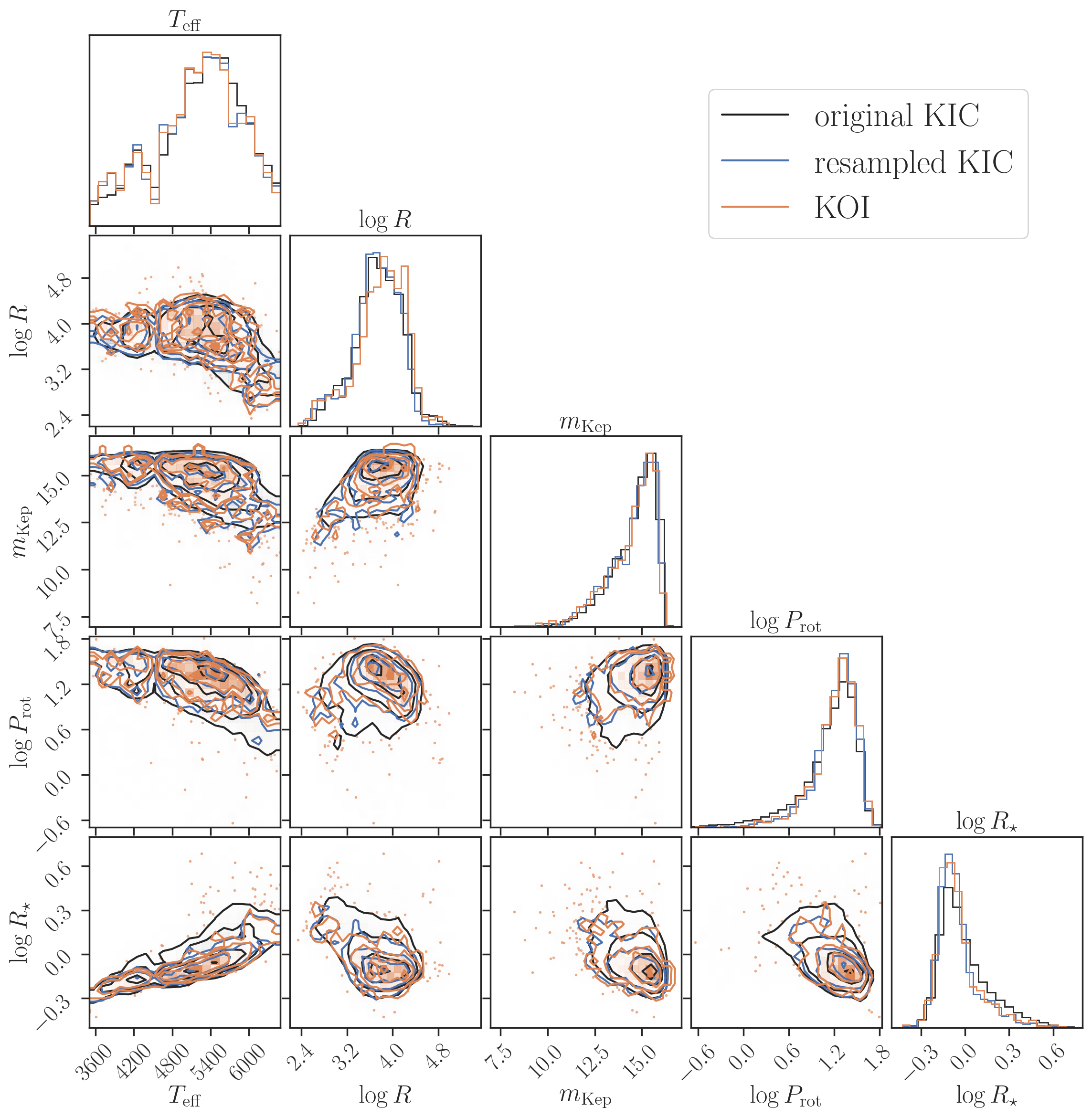}
    \caption{Marginal and pairwise joint distributions of $\teff$, $\amp$, $\mag$, $\prot$, and $\rad$ for the original KIC sample (black), the resampled KIC sample (blue), and the KOI sample (orange). The resampled distribution is represented by 50,000 stars randomly selected from the pooled 200 realizations. The resampling matches $\teff$, $\mag$, $\prot$, and $\rad$ to the KOI sample; $\amp$ is shown for comparison but is not used in the matching. The contours enclose approximately 12\%, 39\%, 68\%, and 86\% of each distribution, corresponding to the default 0.5, 1, 1.5, and 2$\sigma$ levels for a two-dimensional Gaussian distribution.}
    \label{fig:resample_corner}
\end{figure*}

\begin{figure*}[ht]
    \plotone{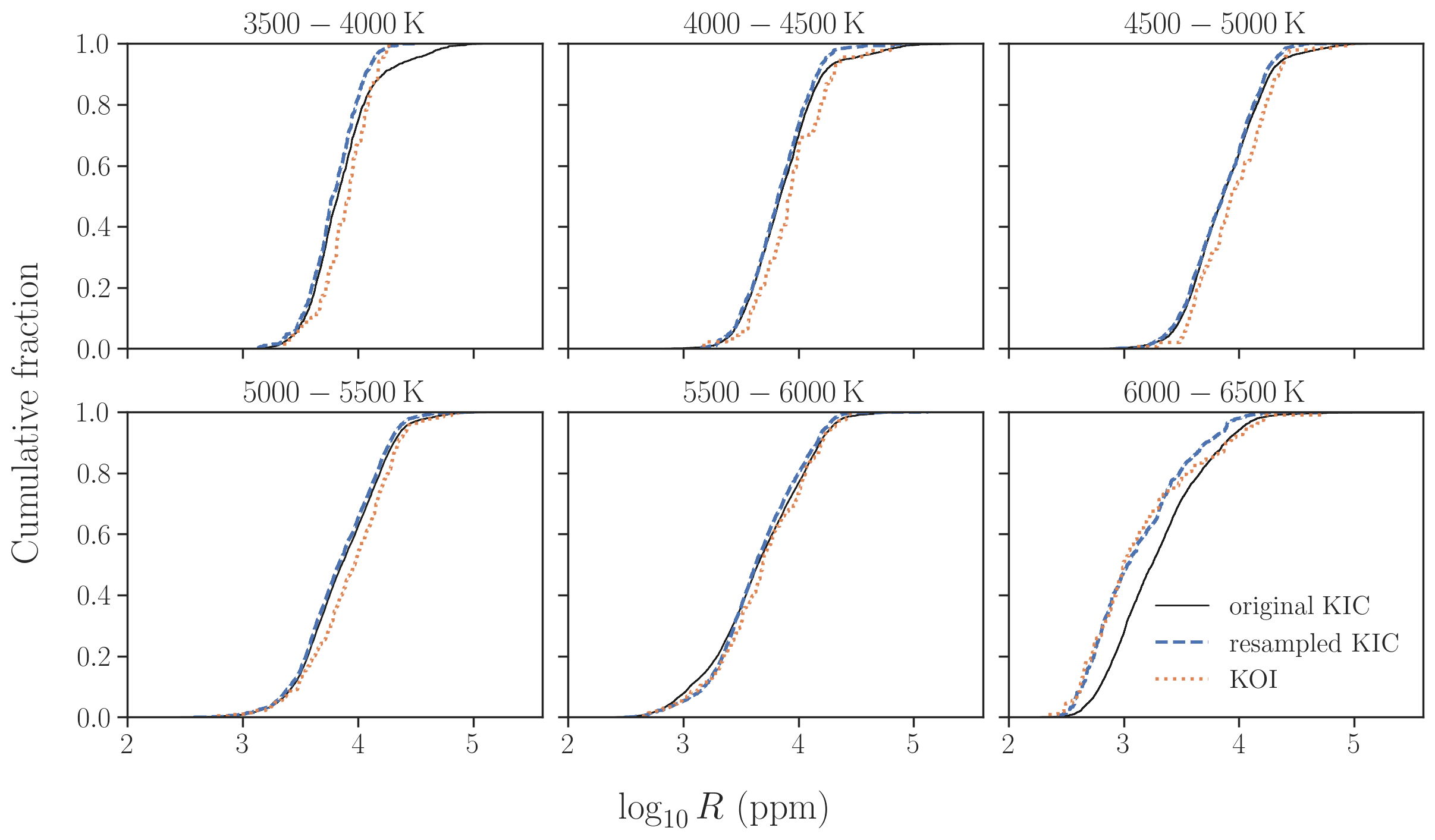}
    \caption{Empirical cumulative distributions of $\amp$ in $500\,\mathrm{K}$ bins for the original KIC sample (black solid lines), the KIC sample pooled over the 200 resampling realizations (blue dashed lines), and the KOI sample (orange dotted lines).}
    \label{fig:resample_ecdf}
\end{figure*}

\subsection{Principle of the Approach}\label{ssec:resampling_principle_normalized}

\newcommand{\pw}{p_\kic^*}

Given the above considerations, we propose to compare the following two samples to examine the difference between the $\amp$ distributions conditioned on $\z$ and $\teff$ (at fixed $\z$ and $\teff$):
\begin{enumerate}
    \item The KOI sample ($\det\tra\rot$) weighted by the narrow $w(\teff-\tau)$ around each $\tau$ and marginalized over the other variables; this is essentially the same as discussed in \citetalias{\mazeh}.
    \item The KIC sample ($\rot$) first weighted by the narrow $w(\teff-\tau)$ and the KOI detectability $\pdet$, and then resampled so that the $\z$ distribution of the detectability-weighted sample at each given $\tau$ aligns with that of the KOI sample above.
\end{enumerate}
The latter sample is designed to subject the KIC sample to the same selection on $\det$ while controlling for variables that could affect the $\amp$ distribution.

The following equations show how these samples meet our requirements. First, the KOI sample defined above in item 1 follows
\begin{align}
    \notag
    p_1(\amp|\tau) &\propto \int p(\amp,\z,\teff|\det\tra\rot)\,\delta(\teff-\tau)\,\d\teff\d\z \\
    &\propto \int p(\amp,\z|\tau,\det\tra\rot)\,\d\z \notag\\
    &= \int p(\amp|\z,\tau,\det\tra\rot)\,p(\z|\tau,\det\tra\rot)\,\d\z \notag\\   
    &= \int {\pdet(\amp,\z,\tau)\,p(\amp|\z,\tau,\tra\rot) \over C_\koi(\z,\tau)}\,p(\z|\tau,\det\tra\rot)\,\d\z, \notag\\
 \label{eq:koi_normalized}
    C_\koi(\z,\tau) &\equiv \int \pdet(\amp,\z,\tau)\,p(\amp|\z,\tau,\tra\rot)\,\d\amp.
\end{align}
On the other hand, the detectability-weighted KIC samples in item 2 follow $\pw(\amp,\z,\teff) \propto \pdet(\amp,\z,\teff)\,p(\amp,\z,\teff|\rot)$, which gives
\begin{align}
    \pw(\amp|\z,\teff) &= {\pdet(\amp,\z,\teff)\,p(\amp|\z,\teff,\rot) \over C_\kic(\z,\teff)}, \quad \notag\\
    C_\kic(\z,\teff) &\equiv \int \pdet(\amp,\z,\teff)\,p(\amp|\z,\teff,\rot)\,\d\amp \notag
\end{align}
at fixed $(\z,\teff)$.
The $\z$ matching is then applied to this weighted distribution to obtain the reweighted and resampled KIC distribution as:
\begin{align}
    \notag
    p_2(\amp|\tau) &\propto \int \pw(\amp,\z,\teff)\,\delta(\teff-\tau)\,{p(\z|\teff,\det\tra\rot) \over \pw(\z|\teff)}\,\d\teff\d\z \\
    \notag
    &\propto \int \pw(\amp,\z|\tau)\,{p(\z|\tau,\det\tra\rot) \over \pw(\z|\tau)}\,\d\z \notag\\
    \notag
    &= \int \pw(\amp|\z,\tau)\,p(\z|\tau,\det\tra\rot)\,\d\z \\
    &= \int {\pdet(\amp,\z,\tau)\,p(\amp|\z,\tau,\rot) \over C_\kic(\z,\tau)}\,p(\z|\tau,\det\tra\rot)\,\d\z. 
        \label{eq:resampled_kic}
\end{align}
Equations~\ref{eq:koi_normalized} and \ref{eq:resampled_kic} show that the two distributions differ only through $p(\amp|\z,\tau,\tra\rot)$ and $p(\amp|\z,\tau,\rot)$, including their corresponding selection normalizations $C$: under the null hypothesis that these conditional distributions are equal, $C_\kic(\z,\tau)=C_\koi(\z,\tau)$, and hence $p_1(\amp|\tau)=p_2(\amp|\tau)$.
Thus, subject to the assumptions discussed above, a difference between $p_1$ and $p_2$ can be attributed to the condition $\tra$.

\subsection{Implementation of the Resampling}\label{ssec:resampling_implementation}

We start with the same KIC and KOI samples analyzed by \citetalias{\mazeh}, for which $\teff$, $\prot$, and $\mag$ are available. For the stellar radii, we adopted the values from \citet{2018ApJ...866...99B}. Cross-matching with this catalog and requiring CDPP measurements for the KIC stars (see below) resulted in final samples of $29700$ KIC stars and $880$ KOIs.

We require an estimate of $\pdet$, the detectability of a transiting planet conditional on transit and marginalized over planet parameters. We estimated this for the KIC stars with detected rotational modulation using a synthetic physical catalog generated within the SysSim framework \citep{2019MNRAS.490.4575H,2020AJ....160..276H}. We selected planets satisfying $|b|<1$ and drew a set of 1,000 period--radius pairs from this transiting population. For each KIC star and simulated planet, we evaluated the expected transit signal-to-noise ratio as
\begin{align}
\mathrm{S/N} &= 10^6\left({R_p\over R_\star}\right)^2
\left({T_\mathrm{obs}f_\mathrm{duty}\over P}\right)^{1/2}
    {1\over \mathrm{CDPP}(T_\mathrm{dur})},\notag\\
T_\mathrm{dur} &= 13\,\mathrm{hr}\left({P\over365.25\,\mathrm{day}}\right)^{1/3}
\left({R_\star\over R_\odot}\right)
    \left({M_\star\over M_\odot}\right)^{-1/3}{\pi\over4}.\notag
\end{align}
The factor $\pi/4$ averages the chord-length factor $\sqrt{1-b^2}$ over the uniform impact-parameter distribution conditional on transit ($|b|<1$).
Here $T_\mathrm{obs}$ and $f_\mathrm{duty}$ are the data span and duty cycle. We evaluated the CDPP at $T_\mathrm{dur}$ by interpolating the tabulated values in log--log space; durations outside the tabulated range were assigned the value at the nearest endpoint. We then converted $\mathrm{S/N}$ into a detection probability using the gamma-function completeness curve 
adopted in \citet{2017AJ....154..109F}, based on the Kepler injection--recovery calibration of \citet{2016ApJ...828...99C}. The value of $\pdet$ for each star was obtained by averaging this probability over the 1,000 simulated planets. Note that this is the detectability conditioned on $\tra$: that is, we evaluate the probability that a planet would be detected given that it transits. The dependence on the geometric transit probability itself is treated separately by controlling for the stellar radius.

We then repeated the following procedure 200 times:
\begin{enumerate}
\item Sample $29700$ stars with replacement from $\kic$, with probabilities proportional to the value of $\pdet$ assigned to each star.
\item In each $100\,\mathrm{K}$ bin over $3500\,\mathrm{K}\leq\teff\leq6500\,\mathrm{K}$, match the $\z=(\mag,\rad,\prot)$ distribution to that of $\koi$. For each KOI, we identify the $k=32$ nearest stars in the sample from step 1 using the scaled distance
\begin{align}
d^2 = \left({\Delta\mag\over0.1}\right)^2
    + \left({\Delta\rad\over0.05}\right)^2
    + \left({\Delta\prot\over0.05}\right)^2,\notag
\end{align}
and draw one of them with probability proportional to $\exp(-d^2/2)$. This sampling is performed with replacement and yields the same number of stars as in the KOI sample in each temperature bin.
\item Compute the median $\amp$ in each $150\,\mathrm{K}$ bin of $\teff$ in the sample from step 2.
\end{enumerate}
This yields 200 realizations of the $\amp$--$\teff$ relation corresponding to $p_2$ in Equation~\ref{eq:resampled_kic}. 

\subsection{Results}\label{ssec:resampling_results}

Figure~\ref{fig:resample_m15} shows the median and 16th--84th percentile interval (KIC resamples; blue circles), together with the KOI sample (orange squares). We find that the resampling lowers the control-sample $\amp$ above $\sim 5500\,\mathrm{K}$ (blue vs gray), where the distributions of $\z$, the transit probability, and the transit detectability vary more strongly than for cooler stars. As a result, the KOI sample no longer lies clearly below the KIC sample; on average, it is higher up to $\sim 6000\,\mathrm{K}$. For hotter stars, the two distributions become indistinguishable within the uncertainties.
Figure~\ref{fig:resample_corner} shows that the resampling approximately matches the relevant stellar-parameter distributions. The marginal and joint distributions of $\teff$, $\mag$, $\prot$, and $\rad$ in the resampled KIC sample closely follow those of the KOI sample, whereas the original KIC sample shows noticeable differences, particularly in the pairwise distributions involving $\mag$ and $\rad$.
Figure~\ref{fig:resample_ecdf} compares the empirical cumulative distributions of $\amp$ in $500\,\mathrm{K}$ bins. Below $6000\,\mathrm{K}$, the KOI distributions are generally shifted toward larger amplitudes relative to the resampled KIC distributions. In the $6000$--$6500\,\mathrm{K}$ bin, the original KIC distribution is shifted toward larger amplitudes than the KOI distribution, while the resampled KIC and KOI distributions nearly overlap.

\section{Likelihood-based Approach}\label{sec:flow}

\newcommand{\rkoi}{\amp_\mathrm{KOI}}
\newcommand{\rkic}{\amp_\mathrm{KIC}}
\newcommand{\zkoi}{\z_\mathrm{KOI}}
\newcommand{\rkoii}{\amp_{\mathrm{KOI},i}}
\newcommand{\ykoii}{\y_{\mathrm{KOI},i}}
\newcommand{\y}{y}

In Section~\ref{sec:resampling}, we compared $p(\amp|\z,\teff,\tra\rot)$ and $p(\amp|\z,\teff,\rot)$ after marginalizing both over the same $\z$ distribution. Although this controls for differences in the distribution of $\z$, 
binning in effective temperature and summarizing the resulting marginalized distributions by their medians may reduce sensitivity to differences in the full conditional amplitude distributions. We therefore complement the resampling analysis with a likelihood-based comparison at each star's observed $\y=(\z,\teff)$, evaluating the required conditional likelihood through a model of the joint distribution of $(\amp,\y)$.

\begin{figure*}[ht]
    \plotone{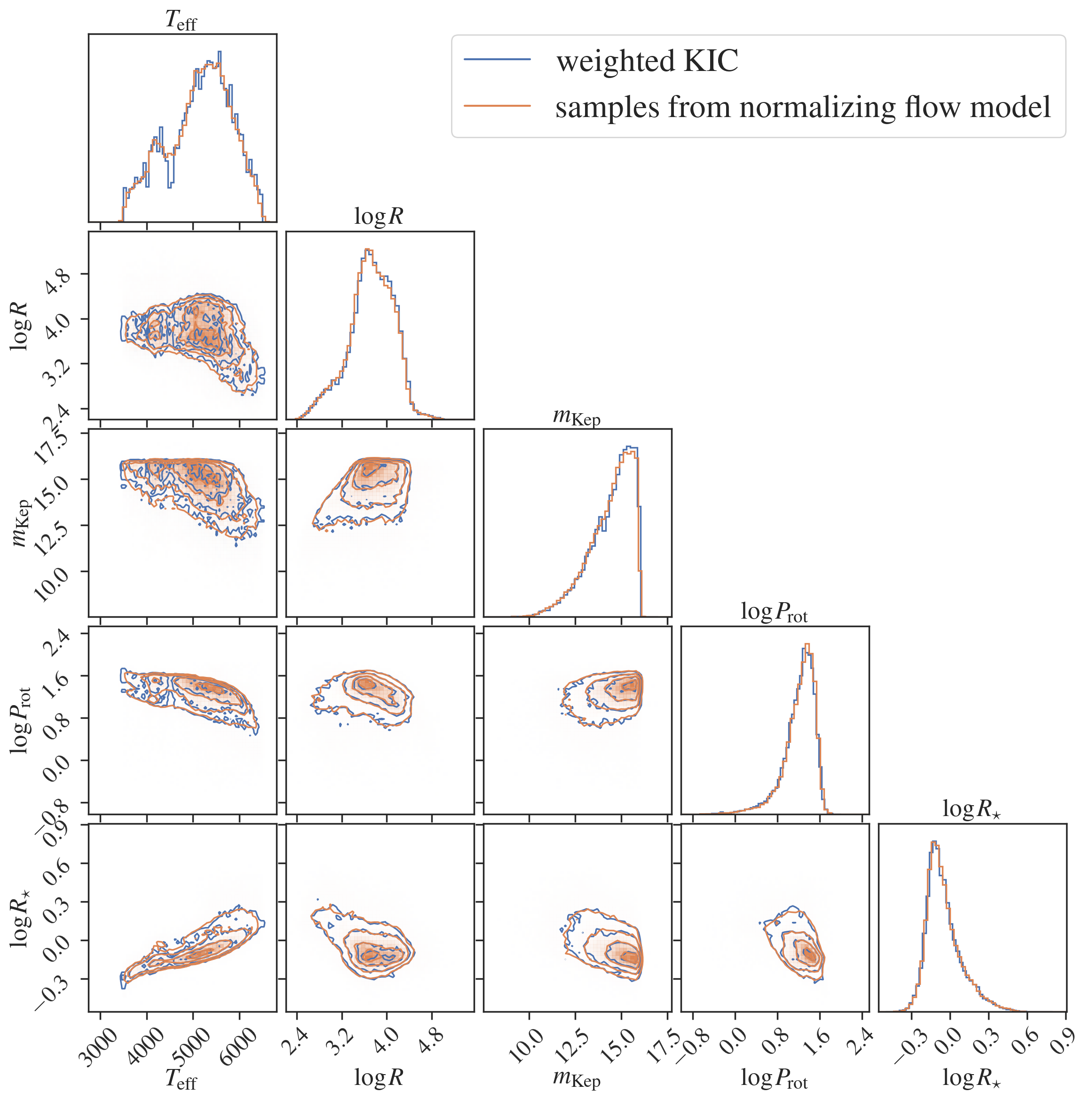}
    \caption{One- and two-dimensional marginal distributions of $(\teff,\amp,\mag,\prot,\rad)$ for the KIC sample weighted by $\pdet$ (blue) and for 60,000 samples drawn from the fitted normalizing-flow model (orange). The diagonal panels show the marginal distributions, and the off-diagonal panels show the joint contours. The contours enclose approximately 12\%, 39\%, 68\%, and 86\% of each distribution, corresponding to the default 0.5, 1, 1.5, and 2$\sigma$ levels for a two-dimensional Gaussian distribution.}
    \label{fig:flow_theta_weighted}
\end{figure*}

\begin{figure*}[ht]
    \plotone{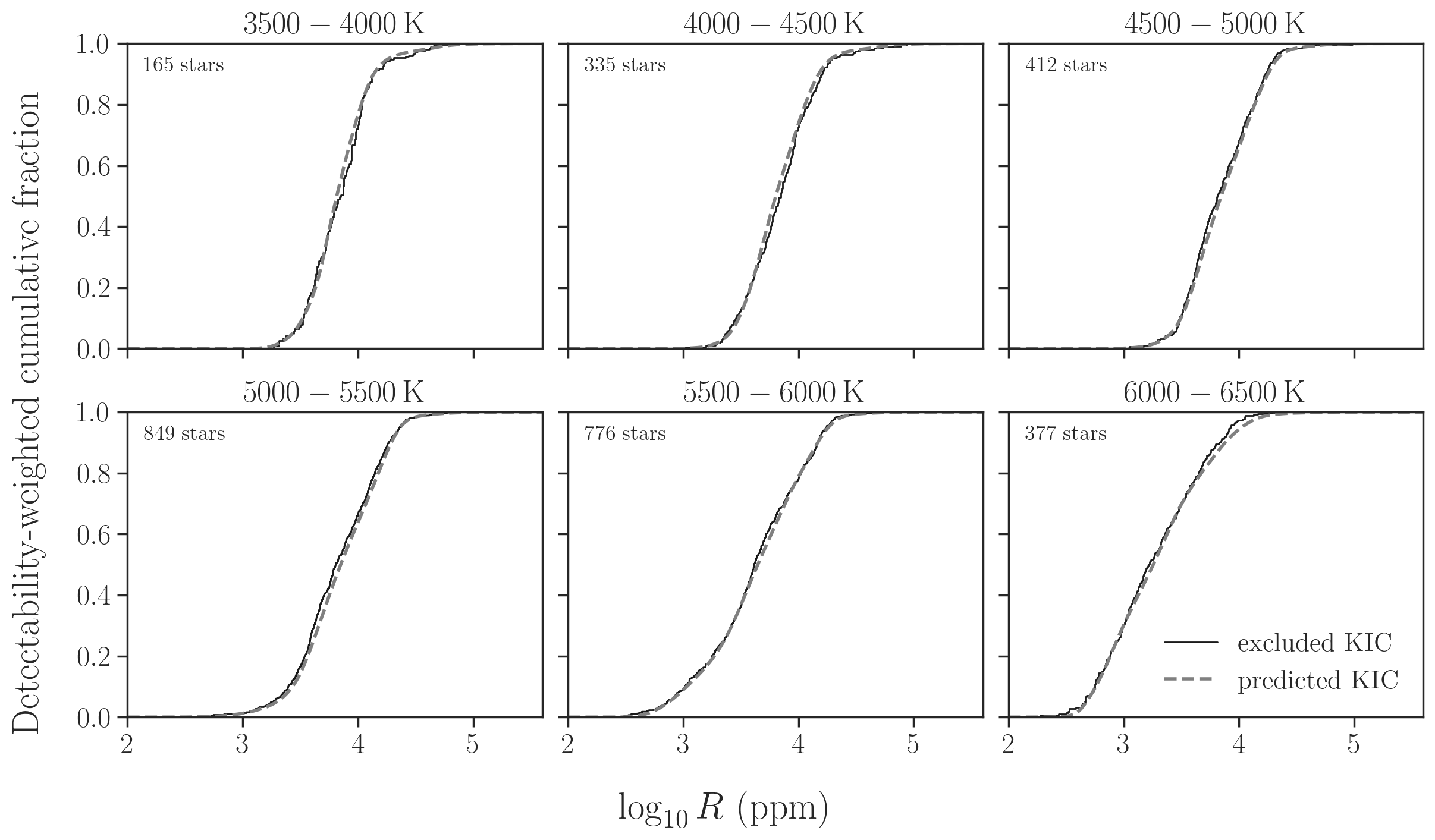}
    \caption{Validation of the flow model using the cumulative distributions of $\amp$ for the 2,914 KIC stars excluded from training. Black solid lines show the empirical cumulative distributions, with each star weighted by $\pdet$. Gray dashed lines show the conditional cumulative distributions predicted at each star's observed $(\teff,\mag,\rad,\prot)$ and averaged over the stars in the panel using the same weights.}
    \label{fig:flow_amplitude_prediction}
\end{figure*}

\begin{figure*}[ht]
    \plotone{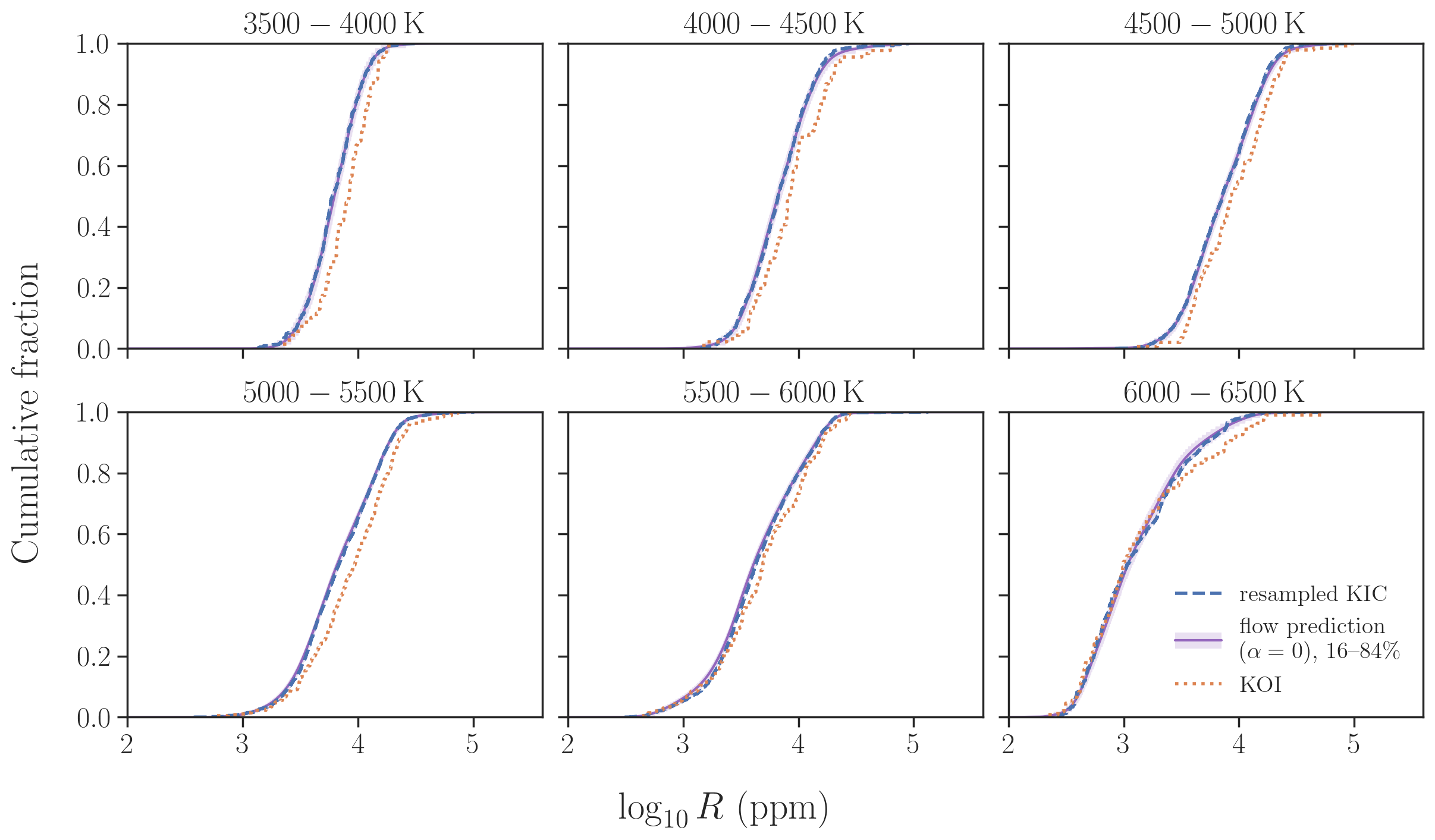}
    \caption{Cumulative distributions of $\amp$ in the same $500\,\mathrm{K}$ bins as Figure~\ref{fig:resample_ecdf}, comparing the KIC sample pooled over 200 resampling realizations (blue dashed lines), the flow predictions under $\alpha=0$ (purple solid lines), and the KOI sample (orange dotted lines). The flow predictions average the conditional cumulative distributions evaluated at each KOI's observed $\y$, with equal weights. The purple bands show the 16th--84th percentile range at each value of $\amp$ among 5,000 simulated KOI samples, generated by independently drawing one amplitude per KOI while holding its $\y$ fixed. These bands describe the variation expected for the observed number of KOIs at fixed flow density and do not include uncertainty in its estimation.}
    \label{fig:koi_flow_prediction}
\end{figure*}

\begin{figure*}[ht]
    \plotone{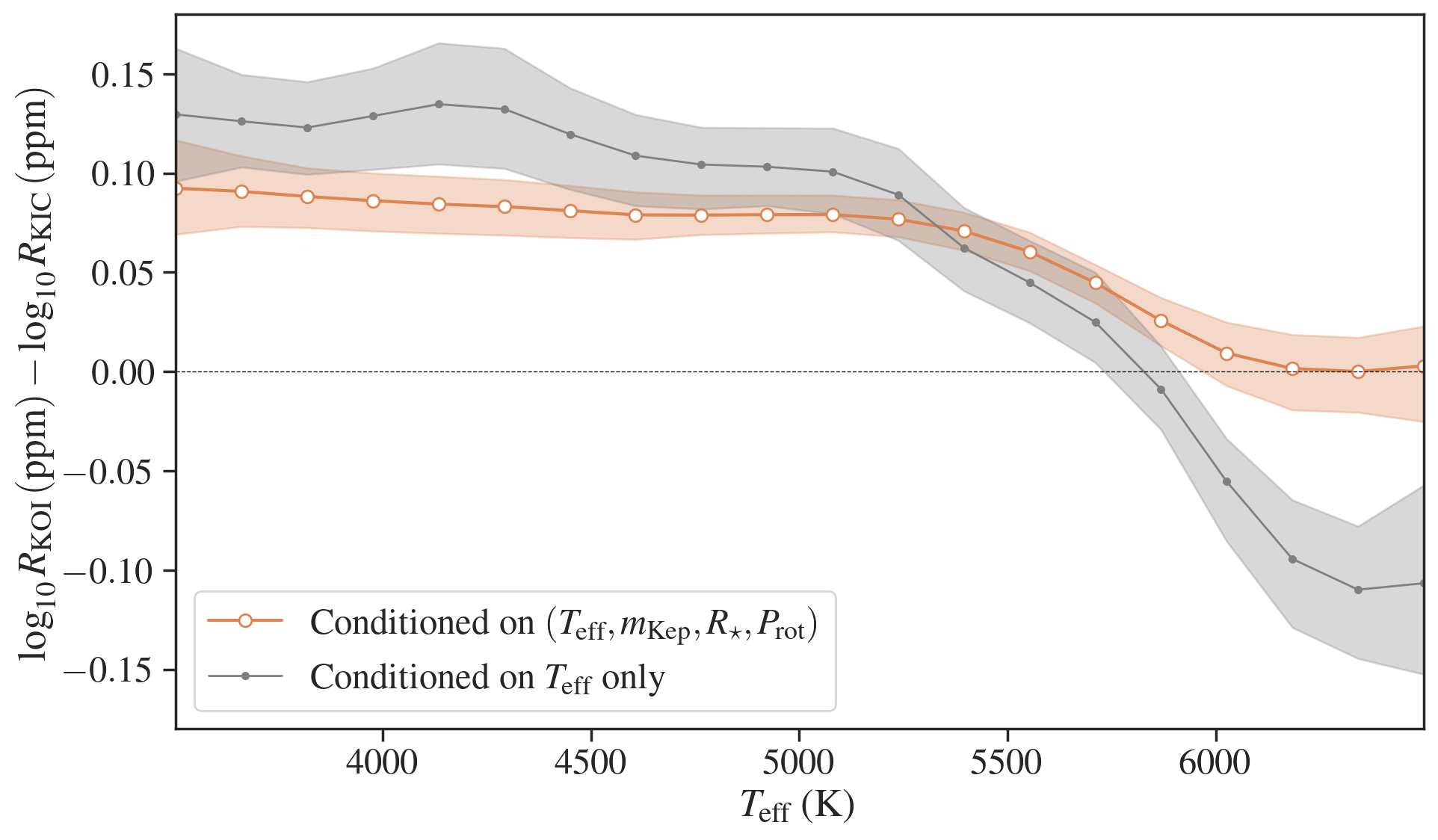}
    \caption{The inferred amplitude shift $\alpha(\teff)$, defined as KOI minus KIC in logarithmic amplitude. The curves and shaded regions show the posterior medians and 16th--84th percentile intervals. The orange curve conditions on $(\teff,\mag,\rad,\prot)$, whereas the gray curve conditions only on $\teff$. Both analyses use KIC reference densities weighted by the detectability of transiting planets, $\pdet$. Positive values indicate larger amplitudes for KOIs.}
    \label{fig:alpha_t}
\end{figure*}

\begin{figure*}[ht]
    \plotone{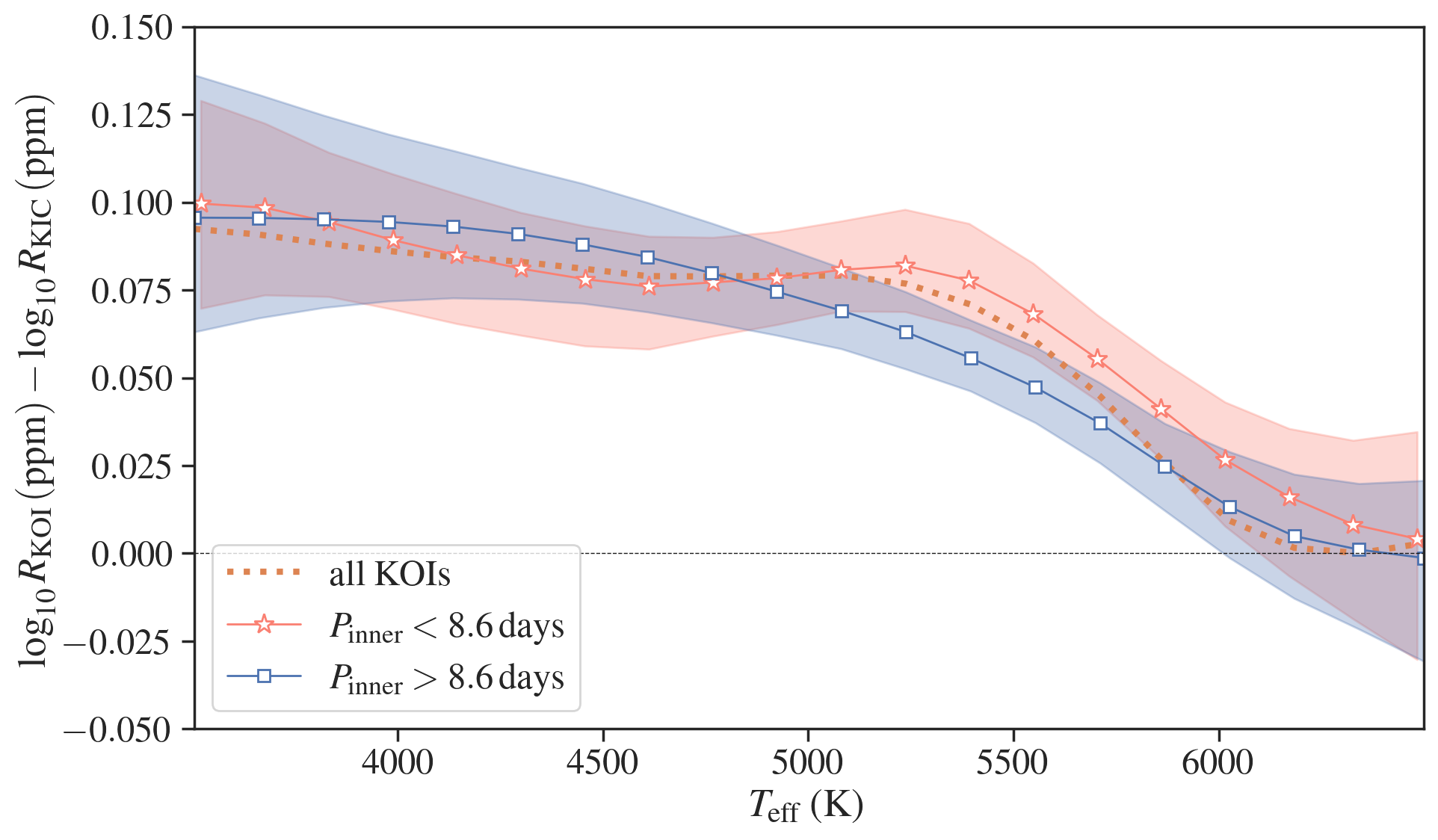}
    \caption{The inferred $\alpha(\teff)$ for all KOIs (orange dotted curve) and for two equal-sized subsamples divided by the orbital period of the innermost planet candidate: $P_{\rm inner}<8.6\,\mathrm{days}$ (salmon stars) and $P_{\rm inner}>8.6\,\mathrm{days}$ (blue squares). The shaded regions show the 16th--84th percentile intervals for the two subsamples. For each subsample, the KIC reference density was trained using $\pdet$ marginalized over the corresponding period range of the simulated planet population.}
    \label{fig:alpha_t_period}
\end{figure*}

\subsection{Conditional Formulation Using a Joint Reference Density}\label{ssec:flow_formulation}

We test whether the distribution of $\rkoi$ at fixed $\y$ is consistent with that of the KIC sample after accounting for planet detectability. The null hypothesis is
\begin{align}
p(\amp|\y,\tra\rot) = p(\amp|\y,\rot) \equiv p_\kic(\amp|\y),\notag
\end{align}
that is, presence of a transiting planet does not affect the $\amp$ distribution at fixed $\y$.
Using Equation~\ref{eq:pdet}, the conditional amplitude distribution expected for KOIs under this null hypothesis is
\begin{align}
p(\amp|\y,\det\tra\rot)
&= {\pdet(\amp,\y)\,p(\amp|\y,\tra\rot)
\over \int \pdet(\amp',\y)\,p(\amp'|\y,\tra\rot)\,\mathrm{d}\amp'} \notag\\
&= {\pdet(\amp,\y)\,p_\kic(\amp|\y)
\over \int \pdet(\amp',\y)\,p_\kic(\amp'|\y)\,\mathrm{d}\amp'} \notag\\
&\equiv p_\kic^*(\amp|\y).\notag
\end{align}

To quantify the departure of the KOI amplitude distribution from this conditional null, we introduce a KOI-minus-KIC shift $\alpha$ in logarithmic amplitude and assume
\begin{align}
\label{eq:conditional_shift_model}
p(\amp|\y,\det\tra\rot)
= p_\kic^*(\amp-\alpha|\y).
\end{align}
Given the observed pairs $(\rkoii,\ykoii)$ and treating $\ykoii$ as fixed, the likelihood function for $\alpha$ is then
\begin{align}
\notag
\mathcal{L}(\alpha)
&=\prod_i p_\kic^*(\rkoii-\alpha|\ykoii) \\
&=\prod_i {p_\kic^*(\rkoii-\alpha,\ykoii)\over p_\kic^*(\ykoii)} \notag\\
&\propto \prod_i p_\kic^*(\rkoii-\alpha,\ykoii), 
\label{eq:conditional_shift_likelihood}
\end{align}
where 
\begin{align}
\label{eq:joint_reference_density}
p_\kic^*(\amp,\y)
\equiv {\pdet(\amp,\y)\,p_\kic(\amp,\y)
\over \int \pdet(\amp',\y')\,p_\kic(\amp',\y')\,\mathrm{d}\amp'\mathrm{d}\y'}.
\end{align}
The last proportionality in Equation~\ref{eq:conditional_shift_likelihood} holds because $p_\kic^*(\ykoii)$ is independent of $\alpha$. 
Thus the inference requires the values of the joint reference density only up to an $\alpha$-independent factor.

We have samples from $p_\kic(\amp,\y)$, rather than its probability density at arbitrary $(\amp,\y)$. In principle, one could construct a multidimensional histogram and evaluate its height to estimate the density. In practice, this becomes difficult because $\y$ contains several parameters, so we instead use the normalizing-flow model described below.

Since $\amp$ is a logarithmic amplitude, $\alpha$ describes a multiplicative shift in the amplitude distribution. This location-shift model is a simplification, although the empirical distributions in Figure~\ref{fig:resample_ecdf} suggest that it provides a reasonable description. The resulting $\alpha$ is only a descriptive parameter for a departure from the detectability-matched, post-selection null distribution $p_\kic^*$. A nonzero value implies that the conditional amplitude distributions of the KOI and KIC samples differ after accounting for planet detectability.
However, the value of $\alpha$ does not necessarily equal an intrinsic shift $\alpha_{\rm int}$ in the underlying pre-selection amplitude distribution, for example one defined by $p(\amp|\y,\tra)=p(\amp-\alpha_{\rm int}|\y)$. Inferring $\alpha_{\rm int}$ would require both the rotational-modulation detection probability $p(\rot|\amp,\y)$, which is not well quantified, and an explicit evaluation of the $\alpha_{\rm int}$-dependent selection normalization. Such an analysis is beyond the scope of this work.

\subsection{Normalizing Flow Model for KIC Stars}\label{ssec:flow_model}

A normalizing flow represents a multidimensional probability density as an invertible transformation of a simple base distribution, such as a multivariate normal distribution \citep{2015arXiv150505770J,2021ITPAM..43.3964K}. The change-of-variables formula then permits direct evaluation of the probability density at any point. A normalizing flow thus provides a smooth and flexible analogue of a multidimensional histogram.
See also \citet{2024AJ....168...80C} for an astronomical application and accessible introduction to normalizing flows.

For the main analysis, we modeled the five-dimensional vector $\bm{\theta}=(\teff,\amp,\prot,\rad,\mag)$. We standardized these variables and constructed a masked autoregressive flow \citep{2017arXiv170507057P} with 16-knot rational-quadratic spline transforms \citep{2019arXiv190604032D} using {\tt FlowJAX} version 17.2.1 \citep{ward2023flowjax}. The model was trained on 29700 KIC stars by maximizing the $\pdet$-weighted log likelihood,
\begin{align}
\widehat{\phi}
=\mathop{\rm arg\,max}_{\phi}
{\sum_n \pdet(\bm{\theta}_n)\log q_\phi(\bm{\theta}_n)
\over \sum_n \pdet(\bm{\theta}_n)},\notag
\end{align}
where $q_\phi$ denotes the flow density. We used a learning rate of $3\times10^{-4}$, a batch size of 1,024, and at most 500 training epochs, with early stopping after 50 epochs without improvement. The fitted density $q_{\widehat{\phi}}$ approximates the detectability-weighted joint reference density $p_\kic^*(\amp,\y)$ in Equation~\ref{eq:joint_reference_density}. It can therefore be evaluated directly at each shifted KOI data point in Equation~\ref{eq:conditional_shift_likelihood}. Figure~\ref{fig:flow_theta_weighted} shows that samples from the fitted flow reproduce the one- and two-dimensional marginal distributions of the weighted KIC sample reasonably well.

With a finite KIC sample, the fitted five-dimensional density need not perfectly represent the underlying reference distribution. As a simple check of the adopted flow model, we randomly excluded 2,914 KIC stars (approximately 10\%) and trained the same flow model on the remaining stars using the same procedure. We then tested whether it could predict the amplitude distribution of the excluded stars. 
For each excluded star, we fixed $\y$ at its observed value and obtained the conditional density $p(\amp|\y)$ by numerically normalizing the fitted joint density over $\amp$, then integrated this density to obtain the cumulative distribution. Figure~\ref{fig:flow_amplitude_prediction} compares the predicted and observed cumulative distributions, weighted by $\pdet$, for these stars. They agree reasonably well within the six temperature bins shown. While this agreement supports the use of the fitted flow as the reference density for the following analysis, this check does not quantify uncertainty in the estimated reference density or its sensitivity to alternative flow architectures or training procedures. The posterior intervals for $\alpha$ reported below are therefore conditional on the fitted reference density. This constitutes a limitation of the present likelihood-based analysis.

The flow fitted to the full KIC sample also provides a model-based counterpart to the resampling comparison in Section~\ref{sec:resampling}. 
We evaluate the conditional distribution $p_\kic^*(\amp|\y)$ at each KOI's observed $\y$ without binning in $\teff$, then average the corresponding cumulative distributions within the same $500\,\mathrm{K}$ bins as Figure~\ref{fig:resample_ecdf} for display. 
The resulting predictions under $\alpha=0$, shown in Figure~\ref{fig:koi_flow_prediction}, essentially correspond to $p_2$ in Equation~\ref{eq:resampled_kic} but averaged over the KOI temperature distribution within each bin. 
They agree reasonably well with the resampled KIC distributions, providing a consistency check between the two approaches. As in the resampling comparison, the observed KOI distributions are generally shifted toward larger amplitudes below $6000\,\mathrm{K}$. Below, we quantify departures from this conditional reference model through $\alpha(\teff)$ using the individual conditional densities.

\subsection{Inference}

We infer $\alpha$ as a function of $\teff$ using the 880 KOIs with all five variables available. We place 20 equally spaced grid points over the temperature range spanned by these KOIs, from $T_{\min}=3{,}500\,\mathrm{K}$ to $T_{\max}=6{,}500\,\mathrm{K}$, and infer the value $\alpha_j$ at each grid point. Linear interpolation between the grid points defines the function $f_\alpha(\teff,\bm{\alpha})$. The likelihood is
\begin{align}
    p(D|\bm{\alpha})
    \propto \prod_i p_\kic^*\!\left(\rkoii-f_\alpha(\teffi,\bm{\alpha}),\ykoii\right),\notag
\end{align}
which we evaluate using the fitted normalizing-flow density $q_{\widehat{\phi}}$. Thus only the amplitude coordinate of each KOI is shifted; its observed $\teff$, $\mag$, $\rad$, and $\prot$ are held fixed.

We place a Gaussian-process prior on $\bm{\alpha}$ to regularize the variation between neighboring grid points. We define the scaled temperature coordinate
\begin{align}
    \widetilde{T} \equiv {\teff-T_{\min}\over T_{\max}-T_{\min}}\notag
\end{align}
and adopt
\begin{align}
\notag
    &p(\bm{\alpha}|\mu_\alpha,\sigma_\alpha,\ell)
    = \mathcal{N}\!\left(\bm{\alpha};\,\mu_\alpha\bm{1},\Sigma_\alpha\right),
    \quad \\
    &(\Sigma_\alpha)_{ij}
    = \sigma_\alpha^2
    \left(1 + {\sqrt{3}\,|\widetilde{T}_i-\widetilde{T}_j|\over\ell}\right)
    \exp\!\left(-{\sqrt{3}\,|\widetilde{T}_i-\widetilde{T}_j|\over\ell}\right),\notag
\end{align}
which is the Mat\'ern-3/2 kernel \citep{2006gpml.book.....R}. Here $\mu_\alpha$ is a constant mean, $\sigma_\alpha$ is the marginal standard deviation, and $\ell$ is a dimensionless correlation length on the scaled temperature coordinate. We restrict each $\alpha_j$ to $[-0.2,0.2]$ and adopt a uniform prior $\mu_\alpha\sim\mathcal{U}(-0.2,0.2)$, along with independent log-uniform priors over $0.005<\sigma_\alpha<0.5$ and $0.05<\ell<5$.

The joint posterior is therefore
\begin{align}
    \label{eq:posterior}
    p(\bm{\alpha},\mu_\alpha,\sigma_\alpha,\ell|D)
    \propto p(D|\bm{\alpha})\,
    p(\bm{\alpha}|\mu_\alpha,\sigma_\alpha,\ell)\,
    p(\mu_\alpha)\,p(\sigma_\alpha)\,p(\ell).
\end{align}
We initialize the chains at a variational maximum-a-posteriori solution and sample Equation~\ref{eq:posterior} using the No-U-Turn sampler \citep{2011arXiv1111.4246H} implemented in {\tt NumPyro} \citep{phan2019composable}. We use four chains, each with 1,000 warm-up steps and 1,000 retained samples, and marginalize over $\mu_\alpha$, $\sigma_\alpha$, and $\ell$. We confirmed that split-rhat \citep{BB13945229} was less than 1.03 for all parameters.

\subsection{Results}

Figure~\ref{fig:alpha_t} shows the inferred $\alpha(\teff)$ in orange. The median shift is approximately $0.07$--$0.09$ over most of the range below $\sim 5{,}500\,\mathrm{K}$, decreases toward higher temperatures, remains positive up to $\sim 6{,}100\,\mathrm{K}$, and becomes consistent with zero for hotter stars.
For comparison, the gray curve in Figure~\ref{fig:alpha_t} shows the same detectability-weighted analysis using only $(\teff,\amp)$ in the joint density, without conditioning on $\z$. This case thus corresponds to the analysis of \citetalias{\mazeh} after accounting for detectability of transiting planets, but not for confounding variables.
This curve crosses zero near the solar temperature and becomes negative at higher temperatures, reproducing the qualitative trend reported by \citetalias{\mazeh}. Because both curves include the same $\pdet$ correction, their difference isolates the effect of conditioning on $(\mag,\rad,\prot)$. Including these variables removes the apparent negative shift for hotter stars and extends the positive KOI--KIC difference to nearly $6,100\,\mathrm{K}$.

This result is qualitatively consistent with the resampling analysis in Section~\ref{sec:resampling}, but the KOI--KIC separation is more pronounced in the likelihood-based analysis. The two approaches need not yield identical offsets because they summarize the conditional difference differently: the resampling analysis compares medians after approximate matching and marginalization over a common $\z$ distribution, whereas the likelihood-based analysis fits a translational shift to the full conditional density at the observed $\z$ of each KOI. The difference may reflect departures from a pure location shift, or the finite-bin and nearest-neighbor approximations used in the resampling procedure.
In addition, as noted in Section~\ref{ssec:flow_model}, the posterior intervals in the likelihood-based analysis do not include uncertainty associated with estimation of the joint density. We therefore place greater emphasis on the qualitative features shared by the two approaches than on the precise magnitude or detailed temperature dependence inferred from the flow-based analysis alone.

Motivated by the period dependence discussed by \citet{2016ApJ...818....5L} and more recently by \citet{2024ApJ...968L...2L}, we also cross-matched the 880 KOIs with the candidate-only KOI catalog used to assign the orbital period of the innermost transiting planet $P_{\rm inner}$ to each star. Of these, 810 had a match; the remaining 70 were excluded from the period-split analysis. We divided the matched KOIs at $P_{\rm inner}=8.6\,\mathrm{days}$, with 405 stars in each subsample. For each subsample, we recomputed $\pdet$ by marginalizing over simulated planets in the corresponding period range and trained a separate weighted KIC flow. Figure~\ref{fig:alpha_t_period} shows that both subsamples follow the same broad temperature dependence as the full sample. The long-period curve declines more gradually, whereas the short-period curve is systematically higher at the sampled points with $\teff\gtrsim 5{,}500\,\mathrm{K}$. This difference is qualitatively consistent with the trend discussed by \citet{2024ApJ...968L...2L}, in which systems with shorter-period planets tend to be more closely aligned. However, the separation is modest, and the result may be sensitive to the accuracy of $\pdet$. We therefore regard the evidence for period dependence as tentative.

\section{Sensitivity to the Adopted Effective Temperatures}\label{sec:teff_robustness}

The effective temperatures adopted by \citetalias{\mazeh} and used in the analyses above are primarily based on the KIC \citep{2011AJ....142..112B}; we refer to them below as the M15 temperatures.
Here we test the sensitivity of our results to the adopted effective-temperature scale by repeating the analysis in Section~\ref{sec:flow} using homogeneous spectroscopic values from the LAMOST DR11 low-resolution AFGK stellar-parameter catalog \citep{2012RAA....12.1197C,2016ApJS..225...28R}.

We constructed LAMOST-matched samples starting from the full catalogs underlying our analysis, which contain 33,614 KIC stars and 993 KOIs. We first associated each KIC identifier with a Gaia DR3 source identifier using the KIC--Gaia correspondence table in \citet{2025A&A...696A.243G},
and then performed an exact match in Gaia DR3 source identifier to the LAMOST DR11 catalog \citep{2026yCat.5162....0L}. 
For a star with multiple LAMOST spectra, we adopted the spectrum having the highest $g$-band signal-to-noise ratio. This procedure yielded finite LAMOST temperatures for 10,779 KIC stars and 338 KOIs.\footnote{Some Kepler stars in the LAMOST DR11 catalog are missed because they lack Gaia DR3 identifiers. However, such stars constitute only about 2\% of the DR11 catalog, so this is unlikely to affect the analysis below.}
For the flow-based analysis, we retained KIC stars with $3500\,\mathrm{K}<T_{\rm eff,LAMOST}<6800\,\mathrm{K}$ for density estimation and KOI stars with $3500\,\mathrm{K}<T_{\rm eff,LAMOST}<6500\,\mathrm{K}$. The wider training interval for the KIC stars provides support beyond the upper boundary of the inference interval. The resulting sample contains 9,401 KIC stars and 287 KOIs.

\begin{figure*}[ht]
    \centering
    \includegraphics[width=0.49\textwidth,clip]{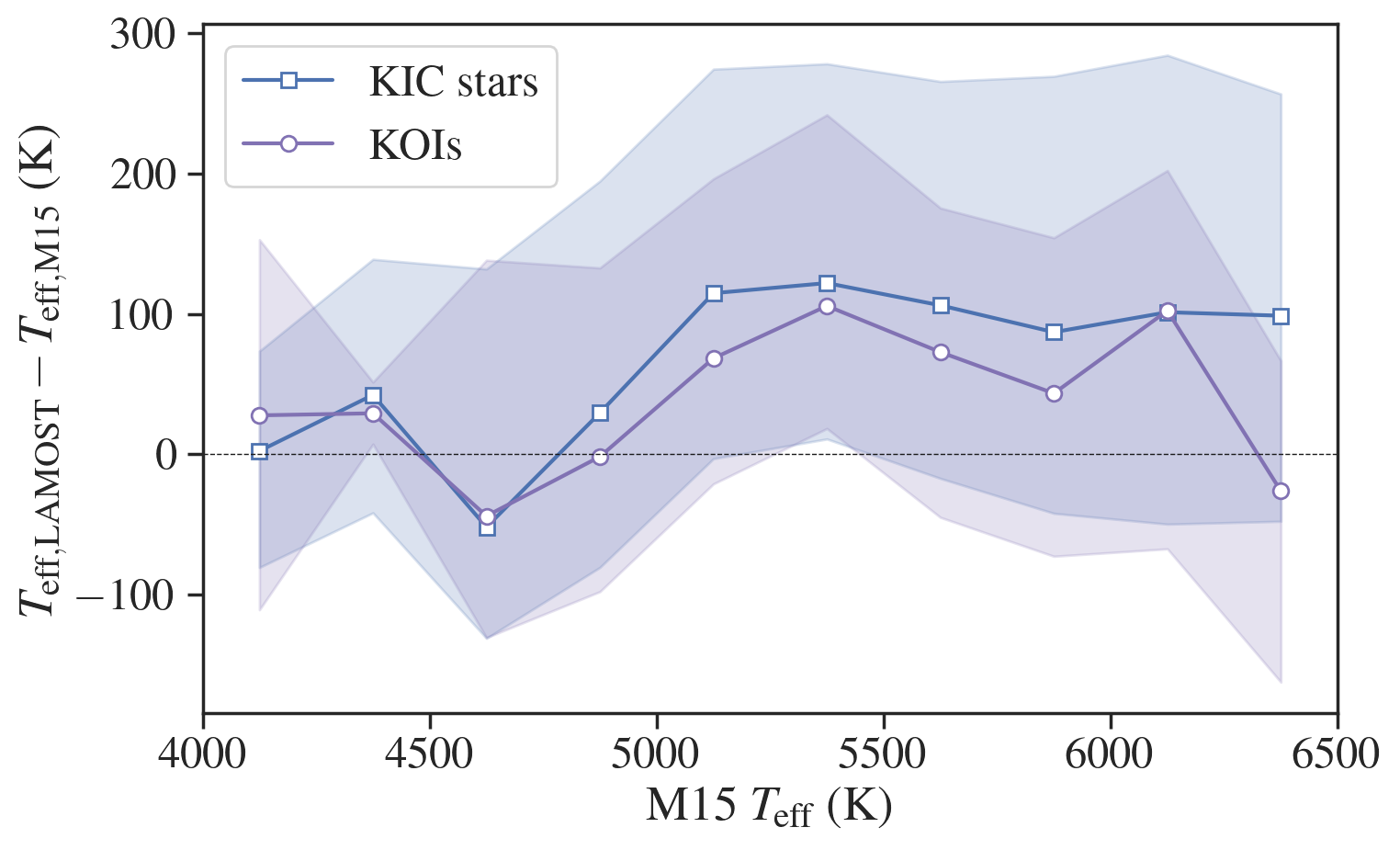}
    \includegraphics[width=0.49\textwidth,clip]{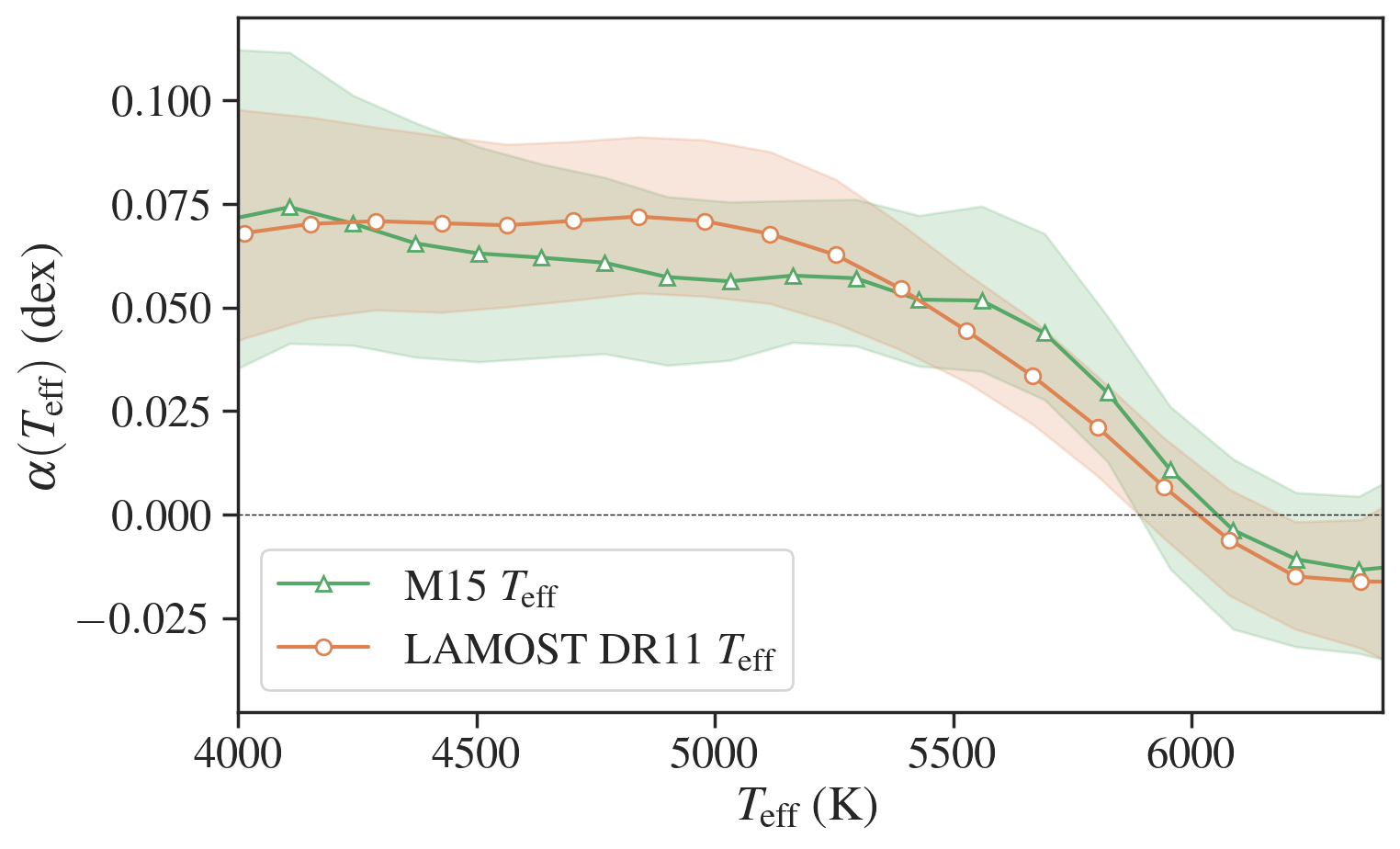}
    \caption{Sensitivity to the adopted effective temperatures. Left: $T_{\rm eff,LAMOST}-T_{\rm eff,M15}$ for the final LAMOST-matched KIC and KOI samples, summarized in $250\,\mathrm{K}$ bins of $T_{\rm eff,M15}$. The curves show the bin medians and the shaded regions enclose the 16th--84th percentiles of the individual temperature differences. Right: the inferred amplitude shift for the same 9,401 KIC stars and 287 KOIs, using either the M15 effective temperatures (green triangles) or the LAMOST DR11 effective temperatures (orange circles) consistently for both the KIC reference density and the KOI likelihood. The curves and shaded regions show the posterior medians and 16th--84th percentile intervals. All inputs and analysis settings other than the effective-temperature measurements are held fixed; only the common, well-supported interval $4000$--$6400\,\mathrm{K}$ is shown in the right panel.}
    \label{fig:alpha_teff_robustness}
\end{figure*}

The two temperature scales broadly agree within $\sim 100\,\mathrm{K}$ (Figure~\ref{fig:alpha_teff_robustness}, left).
The medians and 16th--84th percentile ranges of $T_{\rm eff,LAMOST}-T_{\rm eff,M15}$ are $+88^{+169}_{-129}\,\mathrm{K}$ for the KIC stars and $+54^{+127}_{-123}\,\mathrm{K}$ for the KOIs, respectively. These differences provide a rough indication of the systematic uncertainty associated with the adopted temperature scale.

To further quantify the impact of the adopted temperature scale on the amplitude difference, we repeated the five-dimensional, all-period analysis in Section~\ref{sec:flow} using this sample. We performed two analyses, one with the LAMOST temperatures and the other with the M15 temperatures, while keeping the sample membership fixed to isolate the impact of the temperature measurements from that of selecting the smaller sample. The other four flow coordinates, the $\pdet$ weights, the flow architecture, and the inference model were identical.
The results are shown in the right panel of Figure~\ref{fig:alpha_teff_robustness}.
The inferred curves agree within their posterior uncertainties, suggesting that the analysis is not sensitive to the source of the effective temperatures.
It is worth noting, however, that $\alpha$ is estimated to be smaller than in the full-sample analysis at the highest temperatures, while remaining consistent with zero within the uncertainties.
The same behavior is seen when the M15 temperatures are used, indicating that the difference arises from restricting the analysis to the smaller subset of stars with LAMOST temperatures. This suggests that the sample definition may be more consequential than the choice of temperature catalog.

\section{Discussion}\label{sec:discussion}

\newcommand{\temp}{T_\mathrm{eff}}

\subsection{Interpretation of the Trend}

We find that stars with transiting planets and $\temp\lesssim 6100\,\mathrm{K}$ have larger $\amp$ than stars in the comparison sample. This difference is consistent with a tendency toward low stellar obliquities. 
The inferred shift of $\sim 0.08$~dex is also numerically similar to expectations from models for solar-like distributions of magnetic features. \citet{2020A&A...638A..56N}, for example, found that the mean rotational variability of an isotropically oriented ensemble of otherwise solar-like stars is about 15\% (0.07~dex) lower than that seen equator-on, although we note that the two quantities are not directly equivalent because our $\alpha$ is a descriptive shift of the post-selection distribution and need not equal the intrinsic amplitude shift (as mentioned at the end of Section~\ref{ssec:flow_formulation}).
This interpretation is qualitatively similar to that in \cite{\mazeh}, but after controlling for confounding variables, the photometric evidence consistent with spin--orbit alignment extends to higher temperatures.
Our result provides further support for the low obliquities of small-planet hosts around cool stars inferred from previous stellar-inclination analyses and recent RM measurements (Section~\ref{sec:intro}).

For hotter stars, we find that the $\amp$ distributions of stars with and without transiting planets are consistent with each other. This contrasts with \citetalias{\mazeh}, who found smaller amplitudes among planet hosts, a trend that is not straightforward to interpret geometrically. Thus, the modulation amplitudes no longer provide evidence for an obliquity distribution that would preferentially suppress the amplitudes of planet hosts, such as a predominantly polar distribution \citep[e.g.,][]{2021ApJ...916L...1A,2023AJ....166..112D}.
There are at least three possible explanations for the lack of difference that are not mutually exclusive:
\begin{enumerate}
    \item The two samples have similar inclination distributions because the correlation between stellar spins and planetary orbits in hot-star systems is weak (consistent with a broad obliquity distribution).
    \item The distribution of magnetic features differs between hotter and cooler stars, making $\amp$ less sensitive to the viewing angle for hotter stars.
    \item Conditioning on the detection of rotational modulation ($\rot$) can introduce inclination-dependent selection effects and attenuate the observed star--planet correlation. For example, if rotational modulation in hot stars is detectable mainly for nearly equator-on stars, both samples may be restricted to similar inclination ranges regardless of their underlying obliquity distributions.
\end{enumerate}
Scenarios 2 and 3 imply that modulation amplitudes provide little leverage on the obliquities of these hot stars. 
One motivation for considering scenarios 2 and 3 is that the KOI--KIC difference in $\amp$ begins to diminish over the same $\temp$ range in which $\amp$ declines in the KIC population as a whole (e.g., Figure~\ref{fig:resample_m15}). This coincidence raises the possibility that the properties of magnetic features, such as their characteristic sizes or spatial distribution, change across this temperature range, potentially reducing the sensitivity of $\amp$ to viewing angle. The decline in $\amp$ could also make the detection of rotational modulation more susceptible to selection effects. The extent to which these mechanisms operate remains unclear. Related effects have been demonstrated in specific models \citep{2021ApJ...908L..21R,2023A&A...672A.138N}, but neither study establishes that they occur in the hot-star population considered here.

Although these scenarios are difficult to distinguish using photometric modulation alone, independent $v\sin i$ analyses, which do not rely on spot distributions or the detection of rotational modulation, are consistent with a broad obliquity distribution among hot small-planet hosts and therefore lend support to scenario 1 for these stars \citep{2021AJ....161...68L,2024ApJ...968L...2L}. 
The tentative period dependence in our results may also be qualitatively consistent with that reported by \citet{2024ApJ...968L...2L}. 
Taken together, these results are consistent with a temperature-dependent picture in which small-planet hosts tend to have lower obliquities at lower temperatures and a broader obliquity distribution at higher temperatures. 
The similarity of this trend to that observed for hot Jupiters suggests that the temperature dependence may not arise solely from tidal realignment by hot Jupiters.

The possible similarity in the temperature dependence does not necessarily imply the same transition shape.
Our $\rkoi$--$\rkic$ comparison may instead suggest a relatively smooth change over $500\,\mathrm{K}$, although its inferred smoothness may also be influenced by the Gaussian-process regularization. 
Such a smooth transition is not obviously inconsistent with the $v\sin i$ results of \citet{2021AJ....161...68L,2024ApJ...968L...2L}, who did not attempt to constrain the detailed shape of the temperature dependence.
For hot Jupiters, the relatively sharp transition has motivated explanations involving tidal realignment \citep{2010ApJ...718L.145W,2012ApJ...757...18A,2014ApJ...790L..31D}. The absence of a comparably sharp transition for small planets, even if real, would not be surprising, given their much weaker tidal interactions with their host stars.
However, scenarios 2 and 3 could also contribute to the gradual decline, so the smooth temperature dependence may not be entirely due to a change in the obliquity distribution. Similar $v\sin i$ comparisons spanning this temperature range could help distinguish these possibilities.

\subsection{Considerations for More Accurate Comparisons}\label{ssec:caveats}

Although our analysis accounts for the main identified confounding variables, several aspects could be improved. Because the methodology presented here can also be applied to other analyses, we summarize caveats relevant to future applications.

\paragraph{Have we controlled all the relevant variables?}

Our adopted set of $\z$ approximately controls for stellar evolutionary state and includes variables related to detectability, so it should capture the most important effects.
Nevertheless, it is generally difficult to establish that the choice includes all relevant confounding variables. 
In the present problem, one remaining candidate is metallicity, which affects the thickness of the convective envelope and hence activity and $\amp$, and whose distribution may differ between the KIC and KOI samples through its potential correlation with planet occurrence. If activity amplitude and planet occurrence both increase with metallicity, controlling for it could reduce the inferred KOI--KIC amplitude difference, although the current uncertainties for hot stars are already large. 
Unresolved binary companions may provide another confounder because they can dilute the photometric signal and may also correlate with planet occurrence, particularly for close companions \citep[e.g.,][]{2021MNRAS.507.3593M}.
The flow-based framework presented in Section~\ref{sec:flow} may be useful when $\z$ includes many variables.

\paragraph{Models of the Planet Population and Its Detectability} 

Our comparison avoids the need to model the detection function of rotational modulation, for which we do not have a quantitative model. We nevertheless account empirically for planet detectability using $\pdet$. 
Although $\pdet$ is estimated from injection--recovery simulations performed on actual stellar light curves, it still depends on the unknown underlying planet population over which it is marginalized. 
Thus, limitations of the adopted population model can propagate into $\pdet$ and potentially affect our results.
We adopted a population inferred from the Kepler data, but, for example, did not account for its dependence on $\teff$ or for planet multiplicity \citep[see review by][]{2021ARA&A..59..291Z}.

As a simple sensitivity check, we instead estimated $\pdet$ using the observed KOI population as a function of $\teff$ and found that the inferred trends were not significantly affected.
Although the observed KOI population is itself affected by detection biases and therefore does not represent the true planet population, this agreement suggests that the population assumptions tested here do not strongly affect the results. This likely occurs because the values of $\pdet$ for individual stars are correlated between the two assumptions, so their relative weights change little. 
A more accurate model of the planet population would nevertheless be preferable.

\paragraph{Beyond the Temperature Dependence}

Throughout this paper, we modeled the KOI--KIC amplitude shift as a function of $\teff$. This choice was motivated by the temperature dependence reported for hot Jupiters and by the analysis of \citetalias{\mazeh}, which this work builds on. However, the amplitude shift may depend on other stellar or planetary properties as well. One can instead model $\alpha$ as a function of multiple variables and estimate parameters that quantify the strength of each dependence. This approach would allow possible dependencies to be explored in a data-driven manner. The normalizing flow framework can naturally be extended in this direction.

\paragraph{Model for the Amplitude Difference}

Our model treats $\alpha$ as a translational shift of the KOI $\amp$ distribution relative to the KIC distribution at fixed $\y=(\z,\teff)$, and therefore assumes that the two distributions have the same shape. 
This approximation is sufficient for capturing the amplitude offset as considered by \cite{\mazeh}.
If the difference is driven by stellar inclination, however, the KOI distribution need not be a simple translation of the KIC distribution: a change in the inclination distribution may instead alter its skewness or the relative weight of its high-amplitude tail. A more flexible model could allow both the location and shape of the distribution to vary, or could introduce the stellar inclination as a latent variable and marginalize over its distribution.

A separate limitation concerns the interpretation of this shift. As noted below Equation~\ref{eq:conditional_shift_model}, the inferred $\alpha$ is a descriptive shift away from the null distribution $p_\kic^*$, rather than an intrinsic shift between the pre-selection amplitude distributions of planet hosts and comparison stars. A more precise treatment would therefore require a full forward model of the selection process. In addition to a more accurate model of the planet population and its detectability, as discussed above, such a model would require a quantitative description of the detectability of rotational modulation ($\rot$). 
Such an analysis is possible in principle, but we do not pursue it here because a more detailed statistical model alone would not provide a unique obliquity interpretation: even an intrinsic difference in amplitude distributions could also arise from changes in the distribution of magnetic features on the stellar surface.
Nevertheless, full forward modeling would be a possible future extension once the relevant selection functions and surface-variability models can be calibrated.

\section{Summary and Conclusions}\label{sec:summary}

We revisited the comparison of rotational-modulation amplitudes between Kepler stars with and without detected transiting planets. We showed that a direct comparison is sensitive to differences in stellar properties that are correlated with both sample membership and modulation amplitude, particularly apparent magnitude, stellar radius, and rotation period, as well as to planet detectability. We addressed these effects using two complementary approaches: a resampling analysis that matches the distributions of the relevant stellar properties, and a likelihood-based analysis that enables comparisons at fixed stellar properties by modeling their joint distribution with the amplitude using a normalizing flow.
The agreement between the two approaches indicates that controlling for these effects is particularly important for stars hotter than the Sun.

After these corrections, we find no evidence that stars with transiting planets have smaller modulation amplitudes at any of the temperatures considered. The likelihood-based analysis instead yields a positive shift of approximately $0.07$--$0.09$ dex below $\sim 5{,}500\,\mathrm{K}$; the shift decreases toward higher temperatures, remains positive up to $\sim 6{,}100\,\mathrm{K}$, and is consistent with zero for hotter stars. The resampling analysis gives a qualitatively consistent, although less pronounced, result. Dividing the sample by the period of the innermost planet also hints at a larger shift for shorter-period systems at $\teff\gtrsim 5{,}500\,\mathrm{K}$, but the evidence for this dependence remains tentative.

Provided that modulation amplitude is, on average, positively correlated with stellar inclination, the positive shift for cooler stars can be interpreted as evidence for 
a tendency toward low stellar obliquities. For hotter stars, the lack of a detectable shift is compatible with the broader obliquity distribution inferred from $v\sin i$ analyses and hence with more frequent misalignment. 
The tentative tendency toward larger shifts for shorter-period systems may further suggest that the period dependence of alignment reported by \citet{2024ApJ...968L...2L} extends toward cooler stars.
Taken together, these results are consistent with a possible unified picture in which cool small-planet hosts tend to be aligned and hot hosts are more frequently misaligned.
The qualitative similarity of the temperature dependence to that seen for hot Jupiters suggests that this pattern may not be produced solely by tidal realignment of hot-Jupiter hosts. 
Nevertheless, photometric modulation alone may not provide a unique obliquity diagnostic, because the observed trends could also be affected by the distribution of magnetic features on the stellar surface and by selection effects associated with detecting rotational modulation.
Thus, the obliquity interpretation rests on the combined evidence from photometric modulation and $v\sin i$. 
Future $v\sin i$ measurements and more complete forward models of the relevant selection functions will be important for testing whether the observed temperature dependence reflects a genuine change in obliquity.

More generally, this work demonstrates that comparisons between detected populations require careful treatment of confounding variables and selection effects. The resampling and conditional-density frameworks developed here provide general approaches for such comparisons.

\begin{acknowledgments}

We acknowledge the work of \citet{\mazeh}, which motivated the present study.
We are grateful to the reviewer for their careful reading of the manuscript and constructive suggestions, which helped improve the paper.
We also thank Takato Tokuno for helpful discussions.
Work by KM was supported by JSPS KAKENHI grant Nos.~21H04998, 25K07387, 26K22393, and 26H02074.
The authors used Codex (OpenAI) to assist with language editing and improving the clarity of the manuscript. The authors reviewed all changes and take full responsibility for the content of the manuscript.

\end{acknowledgments}

\software{JAX \citep{jax2018github}, FlowJAX \citep{ward2023flowjax}, NumPyro \citep{phan2019composable}, corner.py \citep{corner}, SysSim \citep{2025ascl.soft07004H}}




\bibliography{references_masuda}
\bibliographystyle{aasjournalv7}



\end{document}